\documentclass[useAMS,usenatbib]{mn2e}
\usepackage{graphicx,amsmath}
\usepackage{amssymb}
\usepackage{natbib}
\usepackage{threeparttable}
\usepackage{microtype}
\usepackage{color}
\def\CIVdblt{{\rm C~}\kern 0.1em{\sc iv}~$\lambda\lambda 1548, 1550$}
\def\MgIIdblt{{\rm Mg~}\kern 0.1em{\sc ii}~$\lambda\lambda 2796, 2803$}
\def\NVdblt{{\rm N}\kern 0.1em{\sc v}~$\lambda\lambda 1238, 1242$}
\def\OVIdblt{{\rm O}\kern 0.1em{\sc vi}~$ 1031, 1037$}
\def\SiIVdblt{{\rm Si~}\kern 0.1em{\sc iv}~$\lambda\lambda1394, 1403$}
\def\AlIIIdblt{{\rm Al~}\kern 0.1em{\sc iii}~$\lambda\lambda1855,1863$}
\def\FeIIdblt{{\rm Fe~}\kern 0.1em{\sc ii}~$\lambda\lambda 2383, 2600$}
\def\NeVIIIdblt{{\rm Ne~}\kern 0.1em{\sc viii}~$ 770, 780$}

\def\NeVIII{\hbox{{\rm Ne~}\kern 0.1em{\sc viii}}}
\def\OI{\hbox{{\rm O~}\kern 0.1em{\sc i}}}
\def\OII{\hbox{{\rm O~}\kern 0.1em{\sc ii}}}
\def\OIII{\hbox{{\rm O~}\kern 0.1em{\sc iii}}}
\def\OIV{\hbox{{\rm O~}\kern 0.1em{\sc iv}}}
\def\OV{\hbox{{\rm O~}\kern 0.1em{\sc v}}}
\def\OVI{\hbox{{\rm O~}\kern 0.1em{\sc vi}}}
\def\OVII{\hbox{{\rm O~}\kern 0.1em{\sc vii}}}
\def\OVIII{\hbox{{\rm O~}\kern 0.1em{\sc viii}}}
\def\NIII{\hbox{{\rm N~}\kern 0.1em{\sc iii}}}
\def\NIV{\hbox{{\rm N~}\kern 0.1em{\sc iv}}}
\def\NVII{\hbox{{\rm N~}\kern 0.1em{\sc vii}}}
\def\CIII{\hbox{{\rm C~}\kern 0.1em{\sc iii}}}
\def\SiIII{\hbox{{\rm Si~}\kern 0.1em{\sc iii}}}
\def\SVI{\hbox{{\rm S~}\kern 0.1em{\sc vi}}}
\def\NeIX{\hbox{{\rm Ne~}\kern 0.1em{\sc ix}}}

\def\AlII{\hbox{{\rm Al~}\kern 0.1em{\sc ii}}}
\def\AlIII{\hbox{{\rm Al~}\kern 0.1em{\sc iii}}}
\def\CaI{\hbox{{\rm Ca}\kern 0.1em{\sc i}}}
\def\CaII{\hbox{{\rm Ca}\kern 0.1em{\sc ii}}}
\def\CrII{\hbox{{\rm Cr}\kern 0.1em{\sc ii}}}
\def\CII{\hbox{{\rm C~}\kern 0.1em{\sc ii}}}
\def\CIII{\hbox{{\rm C~}\kern 0.1em{\sc iii}}}
\def\CIV{\hbox{{\rm C~}\kern 0.1em{\sc iv}}}
\def\CV{\hbox{{\rm C}\kern 0.1em{\sc v}}}
\def\H{\hbox{{\rm H}}}
\def\HI{\hbox{{\rm H~}\kern 0.1em{\sc i}}}
\def\HII{\hbox{{\rm H~}\kern 0.1em{\sc ii}}}
\def\Lya{\hbox{{\rm Ly}\kern 0.1em$\alpha$}}
\def\Lyb{\hbox{{\rm Ly}\kern 0.1em$\beta$}}
\def\Lyg{\hbox{{\rm Ly}\kern 0.1em$\gamma$}}
\def\Lyth{\hbox{{\rm Ly}\kern 0.1em$\theta$}}
\def\Lyfive{\hbox{{\rm Ly}\kern 0.1em$5$}}
\def\Lysix{\hbox{{\rm Ly}\kern 0.1em$6$}}
\def\Lyseven{\hbox{{\rm Ly}\kern 0.1em$7$}}
\def\Lyeight{\hbox{{\rm Ly}\kern 0.1em$8$}}
\def\Lynine{\hbox{{\rm Ly}\kern 0.1em$9$}}
\def\Lyten{\hbox{{\rm Ly}\kern 0.1em$10$}}
\def\HeI{\hbox{{\rm He}\kern 0.1em{\sc i}}}
\def\HeII{\hbox{{\rm He}\kern 0.1em{\sc ii}}}
\def\FeI{\hbox{{\rm Fe~}\kern 0.1em{\sc i}}}
\def\FeII{\hbox{{\rm Fe~}\kern 0.1em{\sc ii}}}
\def\FeIII{\hbox{{\rm Fe~}\kern 0.1em{\sc iii}}}
\def\MnII{\hbox{{\rm Mn}\kern 0.1em{\sc ii}}}
\def\MgI{\hbox{{\rm Mg~}\kern 0.1em{\sc i}}}
\def\MgII{\hbox{{\rm Mg~}\kern 0.1em{\sc ii}}}
\def\MgIII{\hbox{{\rm Mg~}\kern 0.1em{\sc iii}}}
\def\MgIV{\hbox{{\rm Mg~}\kern 0.1em{\sc iv}}}
\def\MgX{\hbox{{\rm Mg~}\kern 0.1em{\sc x}}}
\def\NaI{\hbox{{\rm Na}\kern 0.1em{\sc i}}}
\def\NV{\hbox{{\rm N}\kern 0.1em{\sc v}}}
\def\NII{\hbox{{\rm N}\kern 0.1em{\sc ii}}}
\def\NIII{\hbox{{\rm N}\kern 0.1em{\sc iii}}}
\def\OVI{\hbox{{\rm O}\kern 0.1em{\sc vi}}}
\def\SiII{\hbox{{\rm Si~}\kern 0.1em{\sc ii}}}
\def\SiIII{\hbox{{\rm Si~}\kern 0.1em{\sc iii}}}
\def\SiIV{\hbox{{\rm Si~}\kern 0.1em{\sc iv}}}
\def\SII{\hbox{{\rm S}\kern 0.1em{\sc ii}}}
\def\SIII{\hbox{{\rm S}\kern 0.1em{\sc iii}}}
\def\SIV{\hbox{{\rm S}\kern 0.1em{\sc iv}}}
\def\TiII{\hbox{{\rm Ti}\kern 0.1em{\sc ii}}}
\def\ZnII{\hbox{{\rm Zn}\kern 0.1em{\sc ii}}}
\def\kms{\hbox{km~s$^{-1}$}}
\def\cmsq{\hbox{cm$^{-2}$}}
\def\cc{\hbox{cm$^{-3}$}}
\def\etal{et~al.\ }
\definecolor{red}{rgb}{1.0,0,0}
\newcommand {\apgt} {\ {\raise-.5ex\hbox{$\buildrel>\over\sim$}}\ }
\newcommand {\aplt} {\ {\raise-.5ex\hbox{$\buildrel<\over\sim$}}\ } 

\title[{\OVI}-Broad {\Lya} Probing Warm Gas]{A Pair of {\OVI} \& Broad {\Lya} Absorbers Probing Warm Gas in a Galaxy Group Environment at $z \sim 0.4^{1}$}
\author[Pachat et al]
{
\parbox{\textwidth}{ 
Sachin Pachat$^{2}$ \thanks{E-mail:sachinpc@live.com},  
Anand Narayanan$^{2}$\thanks{Email: anand@iist.ac.in},
Sowgat Muzahid$^{3}$, 
Vikram Khaire$^{4}$,
Raghunathan Srianand$^{4}$, 
Bart P. Wakker$^{5}$,
and Blair D. Savage$^{5}$
} 
\vspace*{10pt}\\ 
$^{1}$Based on observations with the NASA/ESA {\it Hubble Space Telescope}, obtained at the Space Telescope Science \\ Institute, which is operated by the Association of Universities for Research in Astronomy, Inc., under NASA contract NAS 05-26555\\
$^{2}$Indian Institute of Space Science \& Technology, Thiruvananthapuram 695 547, Kerala, INDIA\\  
$^{3}$Department of Astronomy, The Pennsylvania State University, State College, PA 16801, United States.\\  
$^{4}$Inter-University Centre for Astronomy \& Astrophysics, Pune 411007, INDIA.\\  
$^{5}$Department of Astronomy, The University of Wisconsin-Madison, 5534 Sterling Hall, 475 N. Charter Street, Madison WI 53706-1582, USA\\  
}   

\begin{document}

\pagerange{\pageref{firstpage}--\pageref{lastpage}} \pubyear{2015}
\maketitle

\label{firstpage}

\vspace{20 mm}

\begin{abstract}

We report on the detection of two {\OVI} absorbers at $z = 0.41614$ and $0.41950$ (separated in velocity by $|\Delta v| = 710$~{\kms}) towards SBS~$0957+599$.  Both absorbers are multiphase systems tracing substantial reservoirs of warm baryons. The low and intermediate ionization metals in the $z = 0.41614$ absorber are consistent with an origin in photoionized gas. The {\OVI} has a velocity structure different from other metal species. The {\Lya} shows the presence of a broad feature. The line widths for {\OVI} and the broad-{\Lya} suggest $T = 7.1 \times 10^5$~K. This warm medium is probing a baryonic column which is an order of magnitude more than the total hydrogen in the cooler photoionized gas. The second absorber is  detected only in {\HI} and {\OVI}. Here the temperature of $4.6 \times 10^4$~K supports {\OVI} originating in a low-density photoionized gas. A broad component is seen in the {\Lya}, offset from the {\OVI}. The temperature in the broad-{\Lya} is $T \lesssim 2.1 \times 10^5$~K. The absorbers reside in a galaxy overdensity region with 7 spectroscopically identified galaxies within $\sim 10$~Mpc and $\Delta v \sim 1000$~{\kms} of the $z=0.41614$ absorber, and 2 galaxies inside a similar separation from the $z=0.41950$ absorber. The distribution of galaxies relative to the absorbers suggest that the line of sight could be intercepting a large-scale filament connecting galaxy groups, or the extended halo of a sub-$L^*$ galaxy. Though kinematically proximate, the two absorbers reaffirm the diversity in the physical conditions of low redshift {\OVI} systems and the galactic environments they inhabit. 

\end{abstract}


\section{Introduction}\label{sec1}

The detection of circumgalactic and intergalactic warm-hot ionized plasma at low-$z$ ($z \lesssim 0.5$) holds great significance. Models of how galaxies, clusters and structures of larger scales form, make clear predictions that up to $\sim 50$\% of the baryons in the present universe reside in a highly ionized state outside of the stellar environments of galaxies with temperatures in the range of $T \sim 10^5 - 10^7$~K \citep{cen99,dave01,valageas02,smith11,cen13}. The diffuse nature of this gas with densities of $n_{\H} \lesssim 10^{-5}$~{\cc} has precluded its detection via emission in X-rays with the current generation of X-ray detectors. 

Currently, the proven method to trace the baryons in this ionized plasma is to search for their absorption against the spectrum of background quasars. Due to the high temperature of the plasma, the diagnostics commonly pursued are absorption lines from heavily ionized metals such as {\OVI}, {\OVII}, {\OVIII}, {\NeVIII}, and {\MgX} \citep{tripp01,tripp08,savage02,savage05,danforth08,narayanan09,narayanan12}. Compared to {\NeVIII} and {\MgX}, {\OVI} is more ubiquitous because of oxygen's high cosmic abundance. The ion is easily detected even at low $S/N$ through its strong 1032, 1038~{\AA} doublet transitions. Ultraviolet surveys have therefore relied a great deal on {\OVI} as probes to locate undetected large baryonic reservoirs with $N(\H) \sim 10^{20}$~{\cmsq} at low-$z$ \citep{danforth06,danforth08,savage14}. 

Earlier work, both observational and theoretical, has highlighted ambiguities in determining the origin of {\OVI}. Cool ($T \lesssim 10^4$~K) photoionized gas at very low densities ($n_{\H} \sim 10^{-5}$~{\cc}) can produce strong {\OVI} absorption of $N(\OVI) \gtrsim 10^{14}$~{\cmsq} \citep{thom08a,thom08b,tripp08,prochaska11b,muzahid15}. On the other hand, there are instances where {\OVI} seems to be clearly associated with shock-heated warm gas with $T \sim 10^5 - 10^6$~K at both low and high redshifts \citep{savage10,narayanan10a,narayanan10b,savage11a,savage11b,muzahid12,savage14} which harbors a high fraction of the baryons at $z < 0.1$. In complex multiphase absorbers, the distinction between the two possible scenarios of photoionization and collisional ionization is often blurred. 

Predictions from cosmological simulations have also been divided between cold and warm temperatures for the {\OVI} bearing gas \citep{kang05,smith11,tepper11,oppenheimer12}. The differences primarily stem from the cooling calculations implemented in the simulations. As \citet{smith11} and \citet{tepper11} demonstrate, metallicity plays a pivotal role in the temperature of the gas. When chemical abundances are high ($Z \gtrsim 0.1 Z_{\odot}$), excluding the contribution from metals towards the cooling of the plasma can result in a significant overestimation of the temperature. 

The emerging picture is one where, rather than representing a single class of absorber, {\OVI} originates in gas with different physical properties and ionization mechanisms. This affects estimates of the {\OVI} absorber population's contribution to the cosmic baryon inventory. As \citet{tripp08} point out, the $\Omega_b(\Lya)$ from {\Lya} surveys includes the subset of cool photoionized {\OVI}. Thus, a separate estimate of $\Omega_b (\OVI)$ should carefully exclude systems where the {\OVI} comes from photoionized {\Lya} clouds. 

Compared to {\OVI}, the {\OVII} and {\OVIII} ions are conclusive tracers of high temperature plasma. It is difficult to produce these ions via photoionization because it requires photon energies exceeding $138$~eV and $739$~eV respectively. At $T \gtrsim 3 \times 10^5$~K most of the oxygen will be in the {\OVII} state, while at $T \gtrsim 2 \times 10^6$~K it will be in the {\OVIII} ionization state through collisional ionizations. The K-shell transitions of these ions occur at X-ray wavelengths. However, the number of X-ray {\OVII}, {\OVIII} (and {\NeIX}) absorption line detections has been limited. Except for the $z = 0.03$ {\OVII} absorption associated with the Sculptor Wall of galaxies \citep{buote09, fang10}, the robustness of detections of other {\OVII} and {\OVIII} at $z > 0$ have been questioned because of their low significance \citep{cagnoni04, nicastro05, kaastra06}. With the present generation of X-ray spectrographs, it would require integration times of $\sim 100$~Ms to detect these ions with sufficient significance at column densities lower than $\sim 10^{16}$~{\cmsq}, typical of diffuse gas outside of galaxies \citep{yao12}. Thus, for the time being, spectroscopic observations in the ultraviolet remain the most efficient means to probe warm-hot gas at low-$z$. 

In multi-phase intervening absorbers, crucial insights into the ionization of {\OVI} can come from the presence of metal species like {\NeVIII} which is a more reliable tracer of collisionally ionized gas at $T\sim 7\times10^5$ K. Presently, only 7 intervening {\OVI} - {\NeVIII} absorbers are known \citep{savage05,savage11b,narayanan11,narayanan12,meiring13,hussain15}. The sample size is small because of the difficulty in identifying the weak {\NeVIII} doublets against a continuum with low $S/N$ at far-UV wavelengths. In 6 of 7 cases, the constraints set by {\NeVIII} along with {\OVI} clearly established the presence of warm gas with $T \sim 10^5 - 10^6$~K \citep{savage05,savage11b,narayanan09,narayanan11,narayanan12,meiring13}. \citet{hussain15} describe the case where the {\NeVIII}-{\OVI} in an absorber is found to be consistent with an origin in a low density ($n_{\H} \sim 6 \times 10^{-6}$~{\cc}), metal enriched ($\gtrsim Z_{\odot}$) photoionized plasma, with a cloud line of sight thickness of $\sim 186$~kpc. But even in that case, collisional ionization emerges as a viable alternative. 

The information on {\HI} associated with these high ions is central to carrying out a robust measurement of the abundances and the total baryon content \citep{richter06,lehner07,danforth10,danforth11}. The predicted temperatures for the warm absorbers are in the range where the bulk of the hydrogen will be collisionally ionized. The very low neutral fraction of {\HI} ($N(\HI)/N(\H) \sim 10^{-6}$) due to ionization from collisions implies that the {\Lya} absorption from this phase is going to be thermally broadened ($b > 40$~{\kms}), and shallow. Finding this broad-{\Lya} feature (BLA), which is usually has a low contrast with respect to the continuum, requires good $S/N$. Precise line measurements are a challenge especially in multiphase gas, where the shallow BLA is often kinematically entangled with the narrow and typically saturated {\HI} from the cold $T \sim 10^4$~K phase of the absorber. If the BLA is linked in velocity with {\OVI}, it provides a straightforward measurement on the temperature of the gas without having to invoke any underlying modeling assumptions \citep{savage12,savage14}. In these cases, one can use the $b$-parameters of {\HI} and {\OVI} to separate the turbulent and thermal contributions to the line broadening, and thus determine a relatively reliable temperature. 

A complimentary approach towards understanding the origins of {\OVI} absorption has been to investigate the distribution of galaxies at the location of the absorber. Imaging and galaxy spectroscopic surveys have, in several instances, found one or more galaxy counterparts with $L \geq 0.1L^*$ within $\lesssim 300$~kpc impact parameter of {\OVI} absorbers \citep{sembach04,stocke06,tripp06,chen09,wakker09,prochaska11a}. For such low impact parameters, the {\OVI} could be bound to the hot halo of the nearest galaxy. A circumgalactic origin for {\OVI} is also claimed by the more recent COS-Halos survey \citep{tumlinson11a}. This study found a high covering fraction ($\gtrsim 80$\%) for {\OVI} around star forming galaxies, compared to more quiescent systems. The correlation between star formation rate and the incidence of {\OVI} lends support to the notion that the absorption could be part of a galactic wind driven by correlated supernova events within the galaxy. 

Other sets of observations support a slightly different view of the physical locations where {\OVI} bearing gas occurs. From examining the fields of a sample of 20 {\OVI} - BLA absorbers, \citet{stocke14} find that they all occur in galaxy overdensity regions, with a range of impact parameters from $77 - 620$~kpc to the nearest galaxies, with some warm {\OVI} absorbers well outside the virial radius of the nearest galaxy. The temperatures inferred from the {\HI} - {\OVI} line widths were found to correlate with the group's total luminosity as well as the characteristic velocity dispersion of galaxies within the group. This lead them to suggest that the {\OVI} in their sample may be tracing $T \sim 3 \times 10^5$~K gas permeating the space between galaxies, rather than gas within the virial radii of individual galaxies. Similar results, suggesting an intergalactic origin for the {\OVI}, have also emerged from earlier investigations of individual sightlines \citep{narayanan10a,prochaska11b}.
\begin{figure*}
\centering
\includegraphics[totalheight=0.41\textheight, trim=0cm 0cm -1cm 0cm, clip=true]{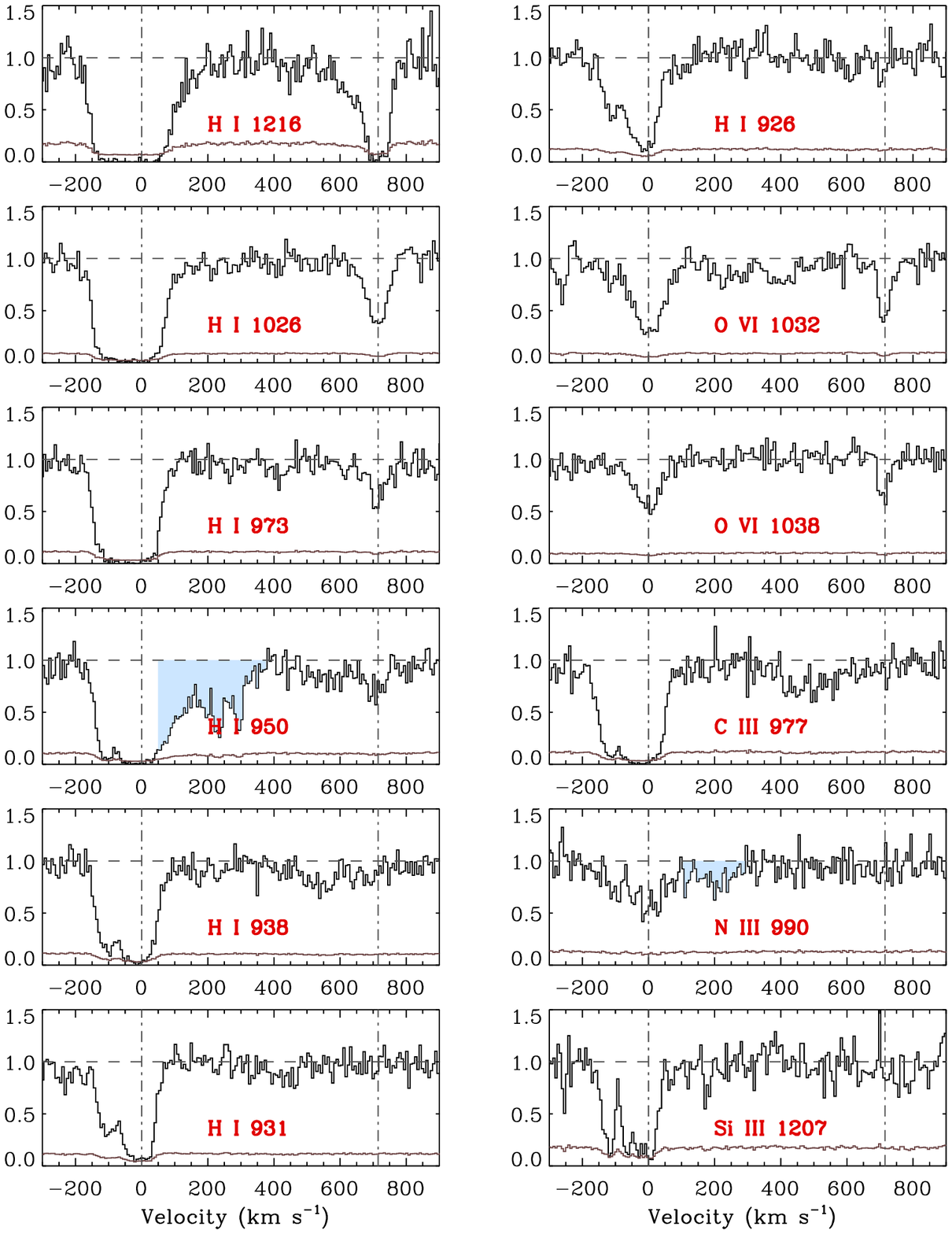}
\includegraphics[totalheight=0.41\textheight, trim=0cm 0cm -1cm 0cm, clip=true]{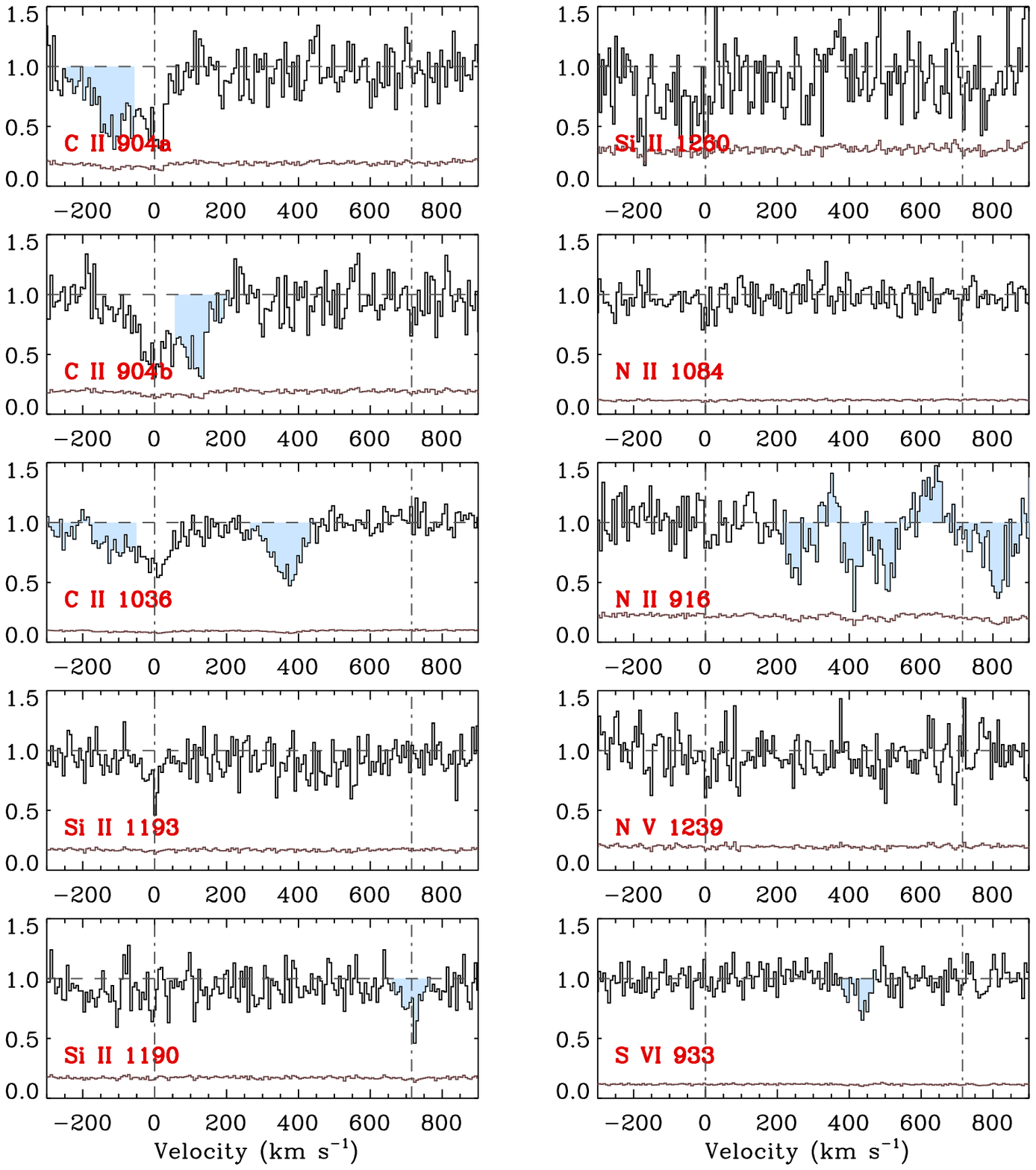}
\caption{Continuum normalized spectral regions of prominent lines associated with the $z = 0.41614$ ($v \sim 0$~{\kms}) and $z = 0.41950$ ($v \sim 710$~{\kms}) absorbers. The vertical lines indicate the relative velocities of the absorbers. For the $v \sim 0$~{\kms} absorber, {\SiII}~1190, {\SiII}~1260, {\NV} and {\SVI} are non-detections at $ > 3\sigma$. Only {\HI}~$1216 - 950$, and {\OVI} are $> 3\sigma$ detections for the $z = 0.41950$ system. The $1\sigma$ error spectrum can also be seen at the bottom of each plot window. The shaded portions indicate absorption unrelated to the line. The absorption contaminating {\HI}~$950$ is identified as {\OVI}~$1032$ at $z = 0.3036$ for which the corresponding {\OVI}~$1038$, {\HI}, {\CIII}, {\SiIII} are also detected. Similarly, the contamination to {\NIII}~$990$ is from interstellar {\SiIV}~$1403$~{\AA}. The {\CII}~904a ($\lambda = 903.9616$~{\AA}) and {\CII}~904b ($\lambda = 903.6235$~{\AA}) are mutually blended at $v < -50$~{\kms} and $v > 50$~{\kms} respectively. Their wavelength separation is only $0.5$~{\AA} in the observed frame. The absorption at $v \sim 400$~{\kms} in the {\CII}~$1036$ panel is {\OVI}~$1038$ from the absorber discussed in this paper. Similarly the feature at $v \sim 710$~{\kms} in the {\SiII}~$1190$ panel is {\SiII}~$1193$ at $v \sim 0$~{\kms}. The absorption features in the {\NII}~$916$ panel from $200 < v < 400$~{\kms} are possibly {\HI}~$916$, {\HI}~$917$, and {\HI}~$918$ lines at $z = 0.4161$. The absorption feature at $v \sim 425$~{\kms} is  {\Lya} at $z = 0.0889$.}
\label{fig1}
\end{figure*}

In the face of these divergent views, there is merit in scrutinizing individual absorption systems in detail, alongside efforts such as the more recent paper by \citet{savage14} that bring out statistical descriptions of {\OVI} absorber populations by surveying a large number of sightlines. Analyzing individual absorbers will help us to describe, with finer detail, the properties of {\OVI} bearing absorbers and the state of diffuse multi-phase gas away from the stellar environments of galaxies. 

In this paper, we report on the detection of two {\OVI} absorbers, kinematically proximate to each other, identified in the $HST$/COS far-UV spectrum of the quasar SBS~$0957+599$. In both absorbers, a BLA is detected, but only in one of them the broad-{\HI} is coincident in velocity with the {\OVI}. The coincidence of the BLA with the {\OVI} allows us to discriminate between a collisional ionization and a photoionization scenario. Whereas, in one of the absorbers, the {\OVI} is tracing $T \sim 7 \times 10^5$~K gas, in the other the {\OVI} is consistent with cooler $T \sim 4.6 \times 10^4$~K photoionized gas. In Sec.~\ref{sec2}, a description of the spectroscopic data is given. Sec.~\ref{sec3}  offers details on both absorbers and the respective line measurements. The ionization mechanisms, absorber physical properties and the chemical abundances are discussed in Secs.~\ref{sec4} \& \ref{sec5}. In Sec.~\ref{sec6}, we provide information on the galaxies detected near the absorber. The possible origin of both multiphase absorbers are dealt with in Sec.~\ref{sec7}, and in Sec.~\ref{sec8} we provide a summary of the significant results.

\begin{figure*}
\centering
\includegraphics[totalheight=0.51\textheight, trim=0cm 0cm -1cm -1.5cm, clip=true]{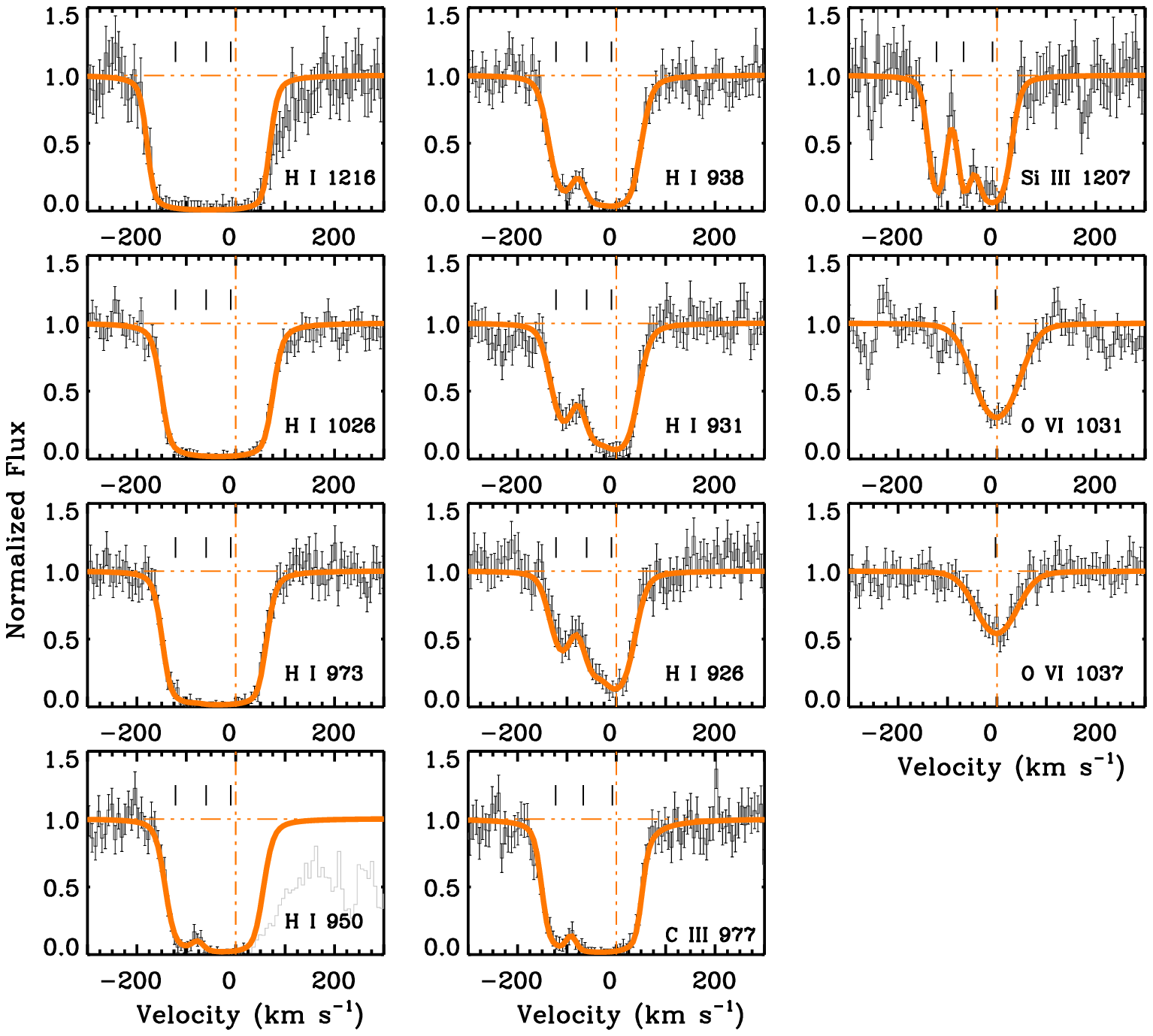}
\caption{Voigt profile models are shown superimposed on the continuum normalized data with $1\sigma$ error bars. The {\HI} absorption is modeled by simultaneously fitting the Lyman series lines. The {\HI} is best fitted with three kinematically distinct components, the positions of which are indicated with the vertical tick marks. The {\HI}~950 {\AA} line is strongly contaminated at $v > 0$~{\kms} by {\OVI}~$1032$ at $z = 0.3036$. While fitting, the contaminated pixels were deweighted by artifically enhacing the noise associated with those pixels. A three component model similar to {\HI} is also obtained from independently fitting the {\CIII} and {\SiIII} lines. In contrast, the {\OVIdblt} are simultaneously fitted with a single component. The kinematic profile of {\OVI} is distinct from {\HI} or the intermediate and low ionization lines of other elements. The higher order Lyman lines, {\CIII} and {\SiIII} at $v \sim -10$~{\kms} are significantly saturated. The fit results are given in Table~\ref{tab1}}.
\label{fig2}
\end{figure*}

\section{The HST/COS Data}\label{sec2}

The ultraviolet $HST$/COS \citep{green12} spectroscopic observations for SBS0957+599 ($z_{em} = 0.746$) were extracted from the MAST public archive\footnote{https://archive.stsci.edu/}. The quasar was observed as part of the COS program to map the gaseous halos of dwarf galaxies (PI. J Tumlinson, Prop ID 12248). The data were reduced using the STScI CalCOS (v3.0) pipeline. The observations consists of far-UV spectra obtained with the G130M and G160M gratings for total integration times of $3.3$~ks and $5.2$~ks respectively. The combined spectrum has a wavelength coverage of $1150 - 1800$~{\AA}, and a wavelength dependent resolution of $R \sim 15,000 - 20,000$ which progressively increases towards longer wavelengths. The COS wavelength calibration has residual errors of $\sim 20$~{\kms}, which are at the level of a resolution element of the instrument \citep{savage11a,meiring13,wakker15}. Within an exposure, the alignment errors are found to vary with wavelength. To reduce the impact of these wavelength dependent offsets between exposures, we cross-correlated ISM and IGM lines between different exposures and then fitted a polynomial to the wavelength-dependent offsets. Afterwards, the centroids of the ISM absorption lines were compared to the centroid of the 21-cm emission in the direction of SBS~$0957+599$ in order to determine a final wavelength scale. The full details of this approach are described in the Appendix of \citet{wakker15}. The COS spectra obtained from the pipeline are oversampled with 6 pixels per resolution element. For our analysis, we rebinned the coadded spectra to the optimal sampling of 2 pixels per resolution element.  The spectrum was normalized to the level of the continuum determined by fitting lower-order polynomials.

\section{The Multiphase Absorbers}\label{sec3}

\subsection{The $z = 0.41614$ Absorber}\label{sec3.1}

This absorber is detected at $ > 3\sigma$ significance in a number of hydrogen Lyman series lines, {\OVI}, {\CIII}, {\NIII}, {\SiIII}, and {\CII}. The continuum normalized line profiles are shown in Figure~\ref{fig1}. The COS spectra also covers {\SiII}, {\NII}, {\SII}, {\NV}, and {\SVI} which yield non-detections. The $z = 0.41614$ redshift is the optical depth weighted center of the {\OVI}~$1032$~{\AA} line. For {\HI}, {\OVI}, {\CIII}, and {\SiIII} we have obtained line parameters through Voigt profile modeling using the \citet{fitzpatrick97} routine. For decomposing the line profiles, the models were convolved with the empirically determined line-spread functions of Kriss (2011) at the observed wavelength of each line. The fit results are shown in Figure~\ref{fig2} and Table~\ref{tab1}. 
\begin{table*} 
\caption{Voigt Profile Measurements} 
\begin{center}  
\begin{tabular}{clrrr}
\hline
$z_{abs}$	&	Transition     &     $v$ (\kms)    	   &       log~$[N~(\cmsq)]$	&	$b$(\kms)     \\   
\hline
\hline
$0.41614$ & {\HI}~$1216 - 926$  & $-122~{\pm}~4$  &   $15.82~{\pm}~0.05$  &   $25~{\pm}~3$     \\
	 &	     		& $-61~{\pm}~6$  &    $15.87~{\pm}~0.31$  &   $21~{\pm}~8$     \\
	 &			& $-10~{\pm}~4$  &    $16.30~{\pm}~0.37$  &   $30~{\pm}~5$     \\  
	 &			& $-3$		  &   $14.02~{\pm}~0.23$  &   $115~{\pm}~7$   \\
\\             
	 & {\OVI}~$1032, 1038$  & $ -3~{\pm}~4$    &   $14.54~{\pm}~0.04$  &   $48~{\pm}~4$    \\
\\
	 & {\CIII}~$977$	& $-120~{\pm}~4$   &   $14.45~{\pm}~0.13$  &   $16~{\pm}~5$     \\
	 &		        & $-67$   	   &   $15.14~{\pm}~0.58$  &   $10~{\pm}~7$     \\
 	 &                      & $-8$   	   &   $16.63~{\pm}~0.63$  &   $20~{\pm}~6$     \\
\\
	 &{\SiIII}~$1207$	& $-123~{\pm}~3$   &   $13.49~{\pm}~0.18$  &   $13~{\pm}~5$     \\
	 & 			& $ -67~{\pm}~5$   &   $13.60~{\pm}~0.25$  &   $9~{\pm}~6$     \\
	 & 			& $  -8~{\pm}~5$   &   $15.17~{\pm}~0.47$  &   $14~{\pm}~8$     \\
\hline
$0.41950$ & {\HI}~$1216 - 926$  & $-3~{\pm}~3$   &   $14.57~{\pm}~0.06$  &   $30~{\pm}~4$  \\
	  &	     		& $-72~{\pm}~5$  &   $13.48~{\pm}~0.42$  &   $58~{\pm}~6$
\\             
	  & {\OVI}~$1032, 1038$ & $0~{\pm}~2$     &   $14.15~{\pm}~0.06$  &   $14~{\pm}~3$  \\
\hline
\hline
\end{tabular}
\end{center}
\small{Comments:~In the $z = 0.41614$ absorber, the multi-component structure is clearly discernible for {\SiIII} and {\HI}. The {\CIII} line is saturated and hence we obtain the $N$ and $b$ values by fixing the centroid of two of its components to match {\SiIII}. We found that for the $v = -8$~{\kms} component, changing the $b$-parameter by a small amount alters the estimated column density significantly indicating that the line is in the flat part of the curve of growth. We adopt a value of $\log N (\CIII) = 16.63^{+0.63}_{-1.10}$ accounting for the uncertainty in column density from the range of $b$-values possible for this component. Moreover, this component is possibly a blend of narrower components unresolved by COS. Similarly, for {\SiIII}, we adopt a column density measurement of $\log N (\SiIII) = 15.17^{+0.47}_{-1.43}$. Compared to the $v \sim -8$~{\kms}, the $v \sim -67$~{\kms} and $v \sim -123$~{\kms} components are not severly saturated. The fit parameters for the broad component in {\HI} were obtained by fixing its velocity to that of {\OVI}. The error in the BLA's $b$-parameter is the statistical error given by the fitting routine. The true error in $b$ will be larger than this estimation as we discuss in Sec~\ref{sec3.1.1}. The Voigt profiles of the separate components in {\HI} are displayed in Figure~\ref{fig3}. The $z = 0.41950$ absorber is detected only in {\HI} and {\OVI}.}
\label{tab1}
\end{table*}

\begin{table*} 
\caption{Apparent Optical Depth Measurements for the $z = 0.41614$ Absorber}
\begin{center}
\small
\begin{tabular}{lccc}
\hline
Line    &     $W_r (m{\AA})$    &    $\log~[N~(\cmsq)]$    &    [-v, +v] (\kms)     \\ 
\hline
{\CII}~$1036$        & $< 60$          &   $< 13.8$             &  [-180, -100]     \\
             & $38~{\pm}~7$    &   $13.57~{\pm}~0.09$  &  [-100, -55]     \\
                     & $107~{\pm}~9$   &   $14.06~{\pm}~0.07$  &  [-55, 50] \\
\\
{\CIII}~$977$        & $208~{\pm}~11$  &   $13.85~{\pm}~0.06$  &  [-200, -85]     \\
                     & $> 456$            &   $> 14.3$            &  [-85, 100]     \\
\\
{\NIII}~$990$        & $43~{\pm}~9$   &   $13.67~{\pm}~0.11$  &  [-180, -100]     \\
             & $42~{\pm}~8$   &   $13.70~{\pm}~0.10$  &  [-100, -55]     \\
             & $106~{\pm}~11$  &   $14.10~{\pm}~0.31$  &  [-55, 40]     \\
\\
{\OVI}~$1032$        & $256~{\pm}~12$  &   $14.49~{\pm}~0.06$  &  [-100, 100]     \\
{\OVI}~$1038$        & $150~{\pm}~13$  &   $14.48~{\pm}~0.05$  &  [-100, 100]     \\
\\
{\SiII}~$1260$         & $< 195$          & $< 13.3$                &  [-180, 50]     \\
{\SiII}~$1193$       & $< 93$           & $< 13.2$                &  [-180, 50]     \\
{\SiII}~$1190$       & $< 90$           & $< 13.4$                &  [-180, 50]     \\
\\
{\SiIII}~$1207$      & $157~{\pm}~14$   & $13.13~{\pm}~0.06$    &  [-180, -100] \\
             & $142~{\pm}~11$   & $13.12~{\pm}~0.07$    &  [-100, -55]     \\
                     & $> 272$       & $ > 13.4 $            &  [-55, 50]     \\
\\
{\NII}~$1084$        & $< 57$           & $< 13.7$              &  [-180, 50]     \\
\\
{\NV}~$1239$         & $< 105$          & $< 13.7$              &  [-180, 50]     \\
\\                     
{\SVI}~$933$         & $< 55$           & $< 13.3$              &  [-180, 50]     \\
\\
\hline
\hline
\end{tabular}
\label{tab2}
\end{center}
\scriptsize{Comments - {\CII}~$904a$ ($\lambda = 903.9616$~{\AA}) \& {\CII}~$904b$ ($\lambda = 903.6235$~{\AA}) are separated by only $\sim 0.5$~{\AA} in the observed frame. The {\CII}~$904b$  at $v > 50$~{\kms} has significant overlap with {\CII}~$904a$. We therefore adopt the measurements done on {\CII}~$1036$. The blue end of {\CII}~$1036$ is contaminated because there is no corresponding absorption seen in {\CII}~$904b$ which is expected to be $20$\% stronger than {\CII}~$1036$. The {\CII} and {\NIII} lines do not show a distinct component structure that can be modeled using Voigt profiles. Hence we have integrated the apparent column density along the velocity interval over which absorption from each component is likely to dominate the contribution from the others. For the lines which are not detected at $\geq 3\sigma$, an upper limit is obtained by integrating over the full velocity range where we expect to find the absorption.} 
\end{table*}

\begin{table*} 
\caption{Apparent Optical Depth Measurements for the $z = 0.41950$ Absorber}
\begin{center}
\small
\begin{tabular}{lccc}
\hline
Line    &     $W_r (m{\AA})$    &    $\log~[N~(\cmsq)]$    &    [-v, +v] (\kms)     \\ 
\hline
{\CII}~$1036$        & $< 31$   &   $< 13.5$  &  [-50, +50] \\
{\CII}~$904b$        & $< 51$   &   $< 13.4$  &  [-50, +50] \\
\\
{\CIII}~$977$        & $< 33$   &   $< 12.8$  &  [-50, +50] \\
\\
{\NIII}~$990$        & $< 36$   &   $< 13.6$  &  [-50, +50] \\
\\
{\OVI}~$1032$        & $88~{\pm}~9$  &   $13.97~{\pm}~0.05$  &  [-50, +50] \\
{\OVI}~$1038$        & $39~{\pm}~10$  &  $13.88~{\pm}~0.11$  &  [-50, +50] \\
\\
{\SiII}~$1193$       & $< 60$  &   $< 12.9$  &  [-50, +50] \\
\\
{\SiIII}~$1207$      & $< 57$  &   $< 12.5$  &  [-50, +50] \\
\\
{\NII}~$1084$        & $< 39$  &   $< 13.6$  &  [-50, +50] \\
\\
{\NV}~$1239$         & $< 69$  &   $< 13.6$  &  [-50, +50] \\
\\                     
{\SVI}~$933$         & $< 33$  &   $< 13.0$  &  [-50, +50] \\
\\
\hline
\hline
\end{tabular}
\label{tab3}
\end{center}
\scriptsize{Comments - All metal lines except {\OVI} are non-detections at the $> 3\sigma$ significance level.} 
\end{table*}


The hydrogen shows multi-component absorption with at least two kinematically distinct components clearly evident. Using the three component absorption profile of {\SiIII} as a guideline, the {\HI} column densities were obtained by simultaneously fitting the Lyman series lines {\HI} $1216 - 926$~{\AA}. The unsaturated higher order Lyman lines provide a unique solution to the {\HI} absorption profile. The total {\HI} column density of $\sim 10^{16.5}$~{\cmsq} obtained from profile fitting indicates that the absorber is only partially optically thick at the Lyman limit. The derived value is comparable within its errors with the {\HI} column density of $\sim 10^{16.3}$~{\cmsq} estimated from the optical-depth at the partial Lyman-limit of $\tau_\mathrm{LL} \sim 0.3$. An accurate measurement of $\tau_{LL}$ requires modeling the full QSO continuum including the higher order Lyman series lines. Here we have attempted a crude estimation of $\tau_{LL}$ by defining a flat continuum around the partial Lyman-limit at $\lambda \sim 1292$~{\AA}.  In addition to the strong absorption in the core, the {\Lya} line shows a broad and shallow absorption in its red wing over the velocity range $[-v, +v] = [70, 200]$~{\kms}. This shallow absorption is not recovered by the three component fit. The significance of this red wing is discussed in Sec~\ref{sec3.1.1}. 

The {\SiIII}~1207~{\AA} line shows a three component profile coinciding in velocity with the {\HI}. However, we note that the component at $\sim -10$~{\kms} can have sub structure to it, which is ambigious at the limited resolution of COS. In the case of {\CIII}, the absorption over the $-90 < v < 100$~{\kms} interval is saturated. We therefore adopt the velocities of the two central components of {\SiIII} to fit the corresponding {\CIII} feature. 

In contrast, the {\OVI} absorption does not show the kinematic complexity seen in {\HI}, {\CIII}, {\SiIII} and {\NIII}. The absorption in either line of the {\OVI} doublet is consistent with a single component. This kinematic difference is suggestive of {\OVI} having a different origin compared to other metal lines. The COS data shows a difference of $7$~{\kms} in the velocity centroid of {\OVI} with the nearest component seen in {\HI}, {\CIII} and {\SiIII}. However, this offset is within the $1\sigma$ uncertainty of the velocity of the model profile, and the wavelength calibration residuals expected for COS spectra. More consequential for the presence of multi-phase is the difference in the kinematics of the absorption of {\OVI} compared with {\HI} and the other metal lines. 

The {\CII}~$904a$ ($\lambda = 903.9616$~{\AA}) is blended with absorption from {\CII}~$904b$ ($\lambda = 903.6235$~{\AA}). The weaker {\CII}~$1036$ profile suggests evidence for multiple components but at low contrast. We note that the feature at $v < -70$~{\kms} in {\CII}~$1036$ is most likely a blend as the corresponding absorption in the stronger {\CII}~$904a$ is not seen. Our fitting routine is unable to converge on a three component fit to the {\CII} data. We therefore resort to the apparent optical depth (AOD) technique of \citet{savage91} to determine the integrated column densities of {\CII} separately for the velocity intervals of the three components seen in {\SiIII}. The details of the apparent column density measurements and equivalent widths are given in Table~\ref{tab2}. The {\NIII}~$990$ line also has component structure that cannot be uniquely identified from profile fitting. We use AOD method to determine the apparent column density for {\NIII} as well. There should be no contamination from {\SiII}~$990$ line in the {\NIII}~$990$ line, since the stronger {\SiII}~$1260$, $1193$, and $1190$ lines are non-detections. 

Our profile fitting analysis shows that the central two components of {\CIII} and {\SiIII} are saturated. A free-fit to the {\SiIII} line gives a value of $\log N(\SiIII) = 15.17~{\pm}~0.47$ and $b(\SiIII) = 14~{\pm}~8$~{\kms} for the strongest component. Assuming a scenario of pure non-thermal line broadening, $b(\SiIII) \sim b(\HI) = 30$~{\kms} for the $v \sim -10$~{\kms} component, we obtain a fit to the line profile with a column density that is $\sim 1.4$~dex smaller. The strong dependency of column density on $b$-parameter implies that the line is in the flat part of the curve of growth. Voigt profile fitting does not yield a unique solution for such saturated lines. The errors returned by the profile fitting routine also do not reflect this uncertainty in column density due to saturation. 

If the {\CIII} and {\SiIII} lines are predominantly non-thermally broadened, then from the better constrained $b(\HI)$, we can estimate lower limits on the column densities for {\CIII} and {\SiIII} that are $1.10$~dex and $1.43$~dex lower than the estimated value.  Alternatively, if the lines are fully thermally broadened, then $b(\CIII) \sim 7$~{\kms} and $b(\SiIII) \sim 5$~{\kms} for the $v \sim -10$~{\kms} component. For these very narrow line widths, the Voigt profile models do not suitably fit the data. From this exercise, we conclude that whereas either component of {\CIII} and {\SiIII} could be as broad as the corresponding {\HI}, they cannot be much narrower than what is seen at the resolution of COS. We therefore adopt a $\log N(\SiIII) = 15.17^{+0.43}_{-1.43}$ and $\log N(\CIII) = 16.63^{+0.63}_{-1.10}$ for the $v \sim -10$~{\kms} component. 

\begin{figure}
\centering
\includegraphics[totalheight=0.35\textheight, trim=0cm 0cm 0cm 0cm, clip=true, angle=90]{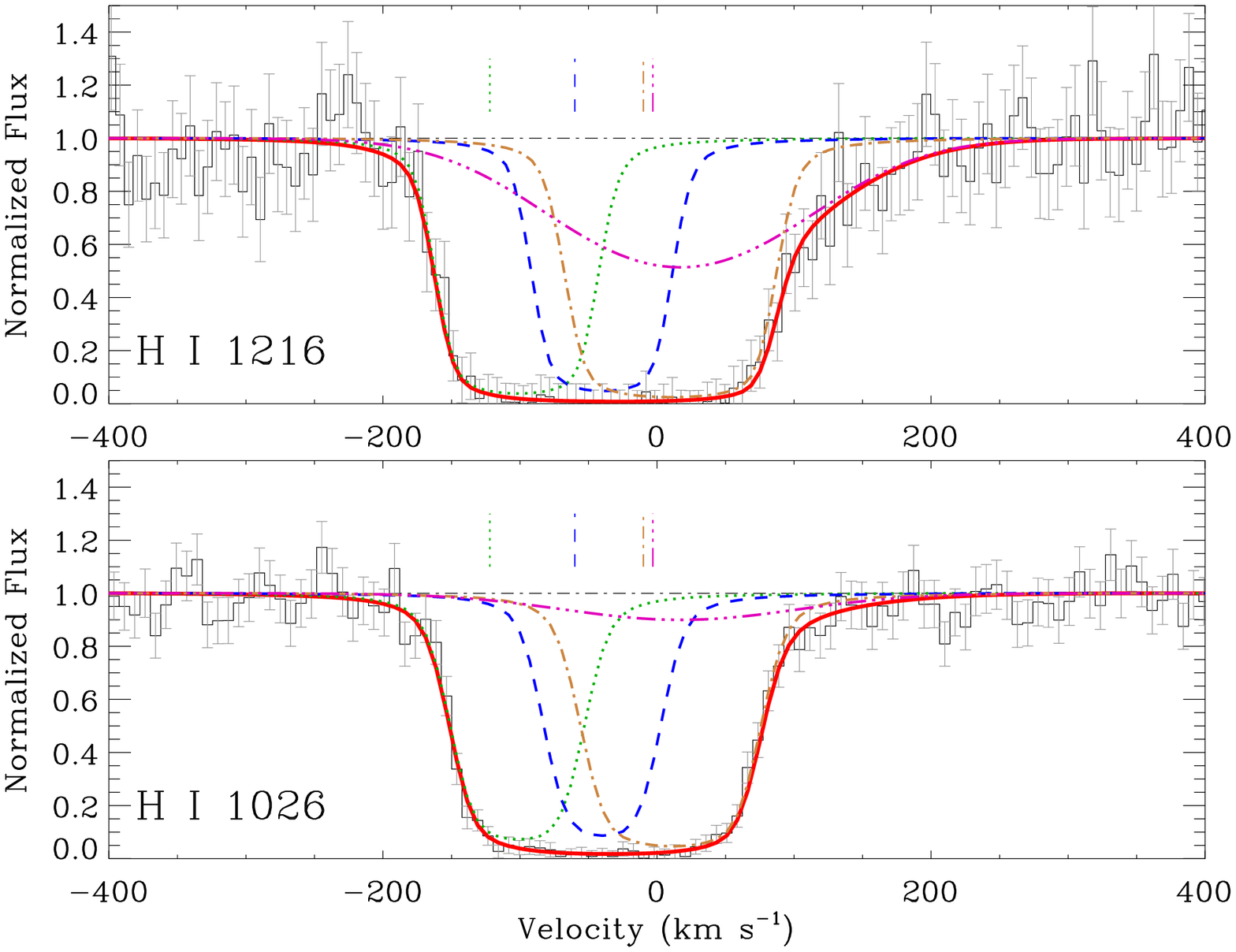}
\caption{Top and bottom panels are Voigt profile fits to {\Lya} and {\Lyb} for the $z = 0.41614$ absorber. The core absorption is simultaneously fitted with three narrow components at $v \sim -122$~{\kms}, $-61$~{\kms}, $\sim -10$~{\kms} shown here as dashed blue curve, dotted green curve and dash-dotted brown curve. The absorption in the red wing of the {\Lya} is fitted by a shallow and broad component with $b \sim 115$~{\kms}. The presence of the BLA is consistent with the {\Lyb} profile where it is not detected. The solid red curve is the combined contribution of all the four components of {\HI}.}
\label{fig3}
\end{figure}
\subsubsection{The Broad Ly-$\alpha$ Absorption}\label{sec3.1.1}

The {\Lya} profile shows excess absorption in the velocity range $[-v, +v] = [70, 200]$~{\kms} over the red wing of the core components. A three component fit to the {\HI} series lines produces an acceptable fit to the strong narrow core components, but does not explain the wing in {\Lya}. To fit this broad feature, we have to introduce a shallow component to the absorption model. By fixing the centroid of this component to the line centroid of {\OVI}, a simultaneous fit to the {\Lya} and {\Lyb} estimates this fourth component of {\HI} to be a BLA with $\log~N(\HI) = 14.02~{\pm}~0.23$~dex and $b(\HI) = 115~{\pm}~7$~{\kms}. The fit results are given in Table~\ref{tab1} and the contributions from the four separate components are shown in Figure~\ref{fig3}. Being a shallow feature, the broad component is not discernible in {\Lyb} or any of the higher order Lyman series lines. Therefore, including these higher order lines to the simultaneous fit does not change the parameters we extract for the BLA. 

The profile fitting routine underestimates the uncertainties in the fit parameters for the BLA. The $1\sigma$ uncertainty in column density of the core {\HI} components, the uncertainties in their $b$-parameters and their line centroids as well as the uncertainty in the $v$ of {\OVI} will influence the fit values the BLA can have. To account for these additional sources of uncertainty in the BLA fit results, we fitted profiles to {\Lya} and {\Lyb} lines simultaneously with the $1\sigma$ uncertainty range of values for $v, N, $~and~$b$ of the core components.  

Combining these deviations from the measured value in quadrature with the statistical uncertainty, we estimate the breadth of the BLA component to be $b(\HI) = 115^{+17}_{-19}$~{\kms}. This is the measurement that we adopt for the BLA for the rest of the analysis. In the case of column density, the statistical error from the Voigt profile fit of $0.23$~dex dominates the uncertainty. The placement of the continuum could also influence the line measurement. But we find that the continuum is well defined within $\Delta v \sim 4000$~{\kms} of the {\Lya}. The large $b$-value implies a temperature of $\log~T$(K) $= 5.91^{+0.12}_{-0.16}$, if the broadening is purely thermal. The BLA reveals the presence of a considerable warm temperature phase to the absorber. 

\subsection{The $z = 0.41950$ absorber}\label{sec3.2}

\begin{figure}
\centering
\includegraphics[totalheight=0.51\textheight, trim=0cm 0cm -1cm -1.5cm, clip=true]{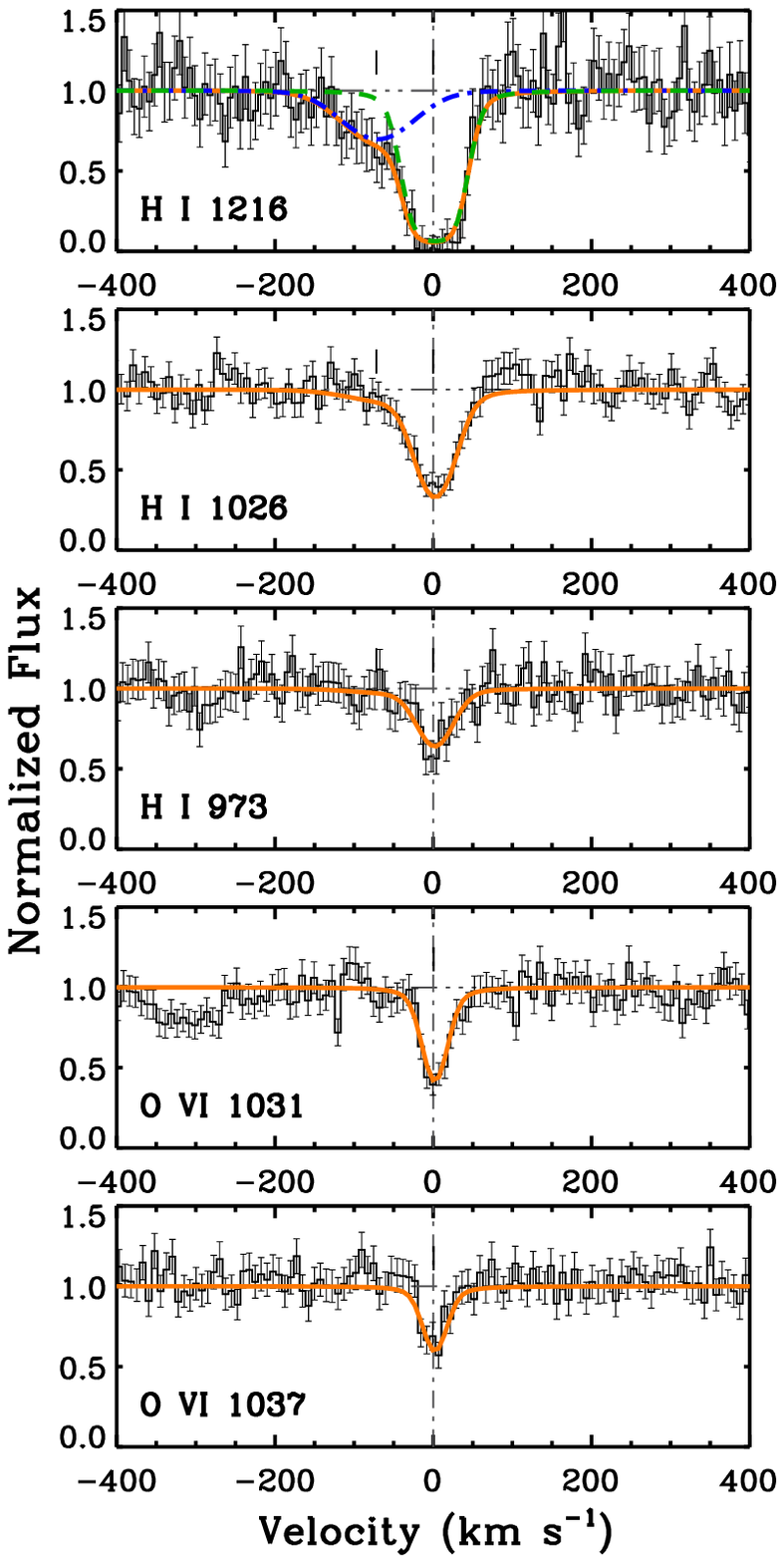}
\caption{Voigt profile models are shown superimposed on the continuum normalized data with $1\sigma$ error bars for the $z = 0.41950$ absorber. The location of the components are indicated by the vertical tick marks. The separate components for {\HI} are shown by the blue dash-dot curve and the green dashed curve. At $v = -72$~{\kms} is a broad component of {\HI} with $b \sim 58$~{\kms} (blue dash-dot line). The orange solid curve superimposed on the Lyman series lines is the composite of the narrow and the broad {\HI} components. As is evident, the broad component's non-detection in the higher order Lyman lines is consistent with its shallow profile seen in {\Lya}. The {\OVI} coincides in velocity with the narrow strong component of {\HI}. The fit results are given in Table~\ref{tab1}.}
\label{fig4}
\end{figure}

The second {\OVI} absorber is offset from the previous system by $\Delta v = +710$~{\kms}. This is a kinematically simple system compared to the $z = 0.41614$ absorber. Only {\HI} and {\OVI} are detected at this redshift with $> 3\sigma$ confidence. Figure~\ref{fig4} shows the profile fit on these lines and Table~\ref{tab1} lists the column density measurements from profile fitting. The upper limits on column densities for the other metal ions are given in Table~\ref{tab3}.  The {\Lya} shows {\HI} absorption in two components. A Voigt profile fit identifies the two components at $v = -3$~{\kms} and $v = -72$~{\kms} with the second component being broader having $b(\HI) = 58$~{\kms} (BLA). The {\OVI} is kinematically coincident with the $v = -3$~{\kms} {\HI} component, as shown by a free fit to the {\OVI} doublet lines. This indicates that the {\OVI} resides in the same gas phase the narrower {\HI} component. The difference in $b$-parameter between {\HI} and {\OVI} implies a kinetic temperature of $T = 4.6^{+0.3}_{-0.2} \times 10^4$~K for this gas phase.

\begin{figure}
\centering
\includegraphics[totalheight=0.27\textheight, trim=0cm 0cm 0cm 0cm, clip=true]{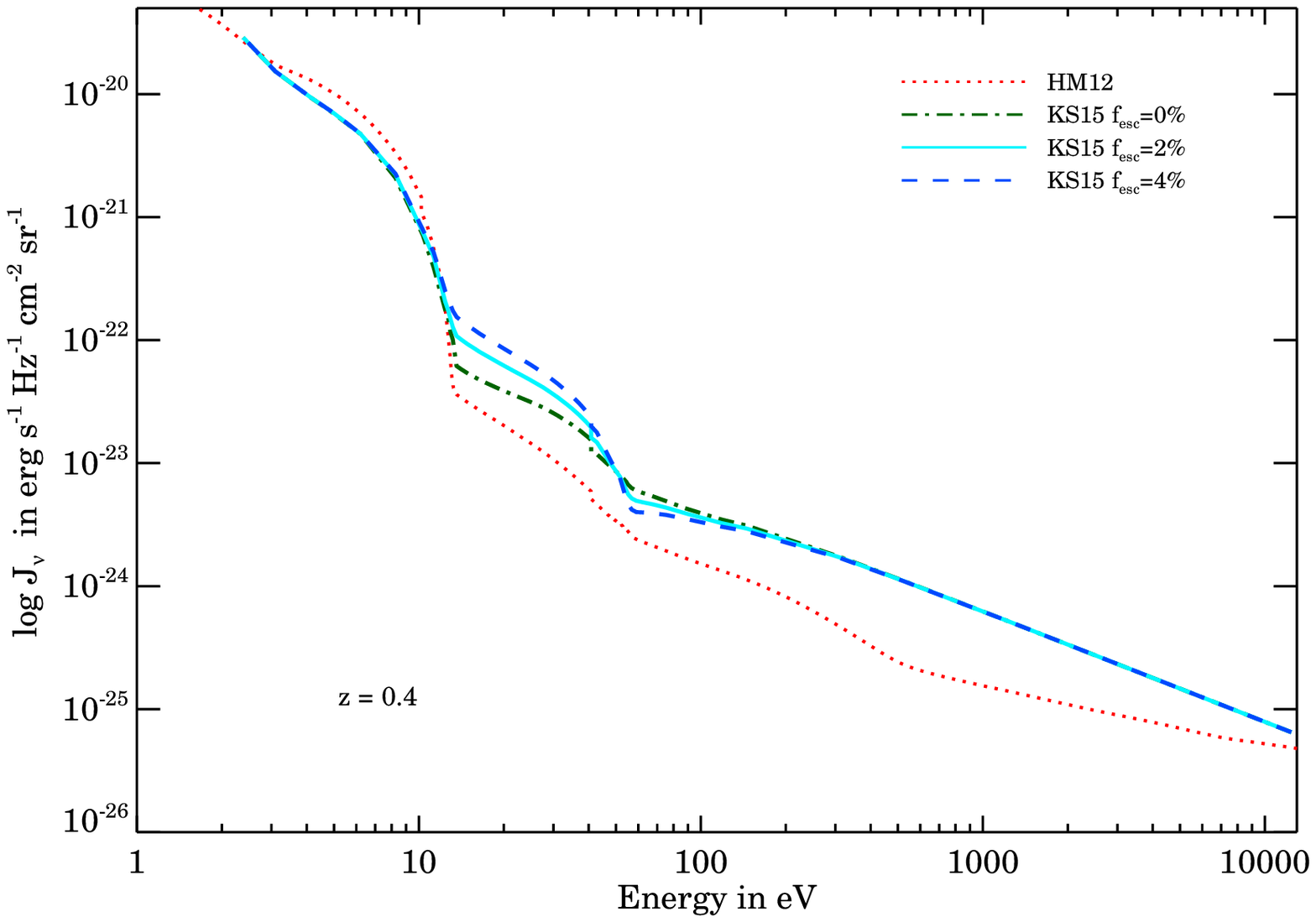}
\caption{The specific intensity of the extragalactic ionizing background radiation for $z = 0.4$. The red dotted curve is the conventional \citet{HM12} ultraviolet background. The other curves are the \citet{KS15b} modeling of the ionizing background which incorporates the more recent measurements on the quasar luminosity function and estimates on escape fraction of Lyman continuum photons from young star forming galaxies. The different \citet{KS15b} curves are for different escape fractions. The trend seen around the {\HeII} ionizing edge ($54.4$~eV), where increasing the escape fraction of photons decreases the intensity of {\HeII} ionizing UVB is explained in \citet{KS13}. Throughout our photoionization modeling, we use the background with $f_{esc} = 4$\%, which will produce the {\HI} photoionization rate required by \citet{kollmeier14} to solve the apparent photon underproduction crisis.}
\label{fig5}
\end{figure}

The broader {\HI} component (BLA) suggests the presence of gas at $T = 2.1 \times 10^5$~K, if the line width is purely thermal. At the velocity of this BLA, no metal line absorption is detected. The BLA feature is not detected in the higher order Lyman lines which is consistent with its shallow profile in {\Lya}. 

\section{Ionization \& Abundances in the $z = 0.41614$ Absorber}\label{sec4}

\subsection{Photoionized Gas Phase}\label{sec4.1}

While the intermediate ions like {\SiIII} and {\CIII} showing evidence for saturation, there are not enough constraints to develop complete ionization models. We draw only general conclusions about the physical state of the absorber from the photoionization models. The photoionized gas in this absorber is modeled using Cloudy \citep[v13.03]{ferland13}. The elemental abundances used are the most recent solar abundances from \citet{grevesse10}. We model the ionization in the $v \sim -122, -61, -10$~{\kms} components of the absorber separately.

\begin{figure*}
\centering6
\includegraphics[totalheight=0.24\textheight, trim=0cm 0cm 0cm 0cm, clip=true, angle=90]{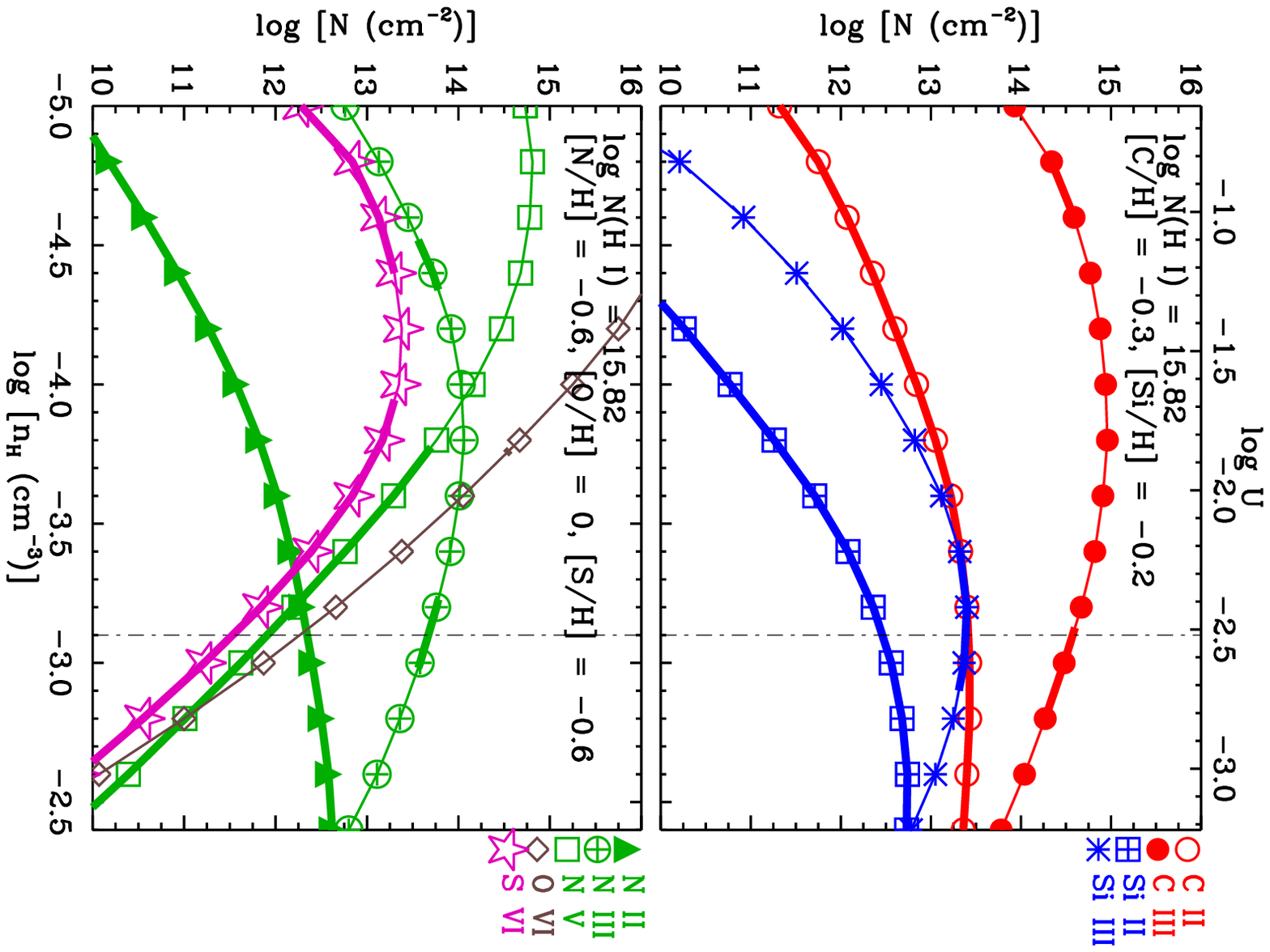}
\includegraphics[totalheight=0.24\textheight, trim=0cm 0cm 0cm 0cm, clip=true, angle=90]{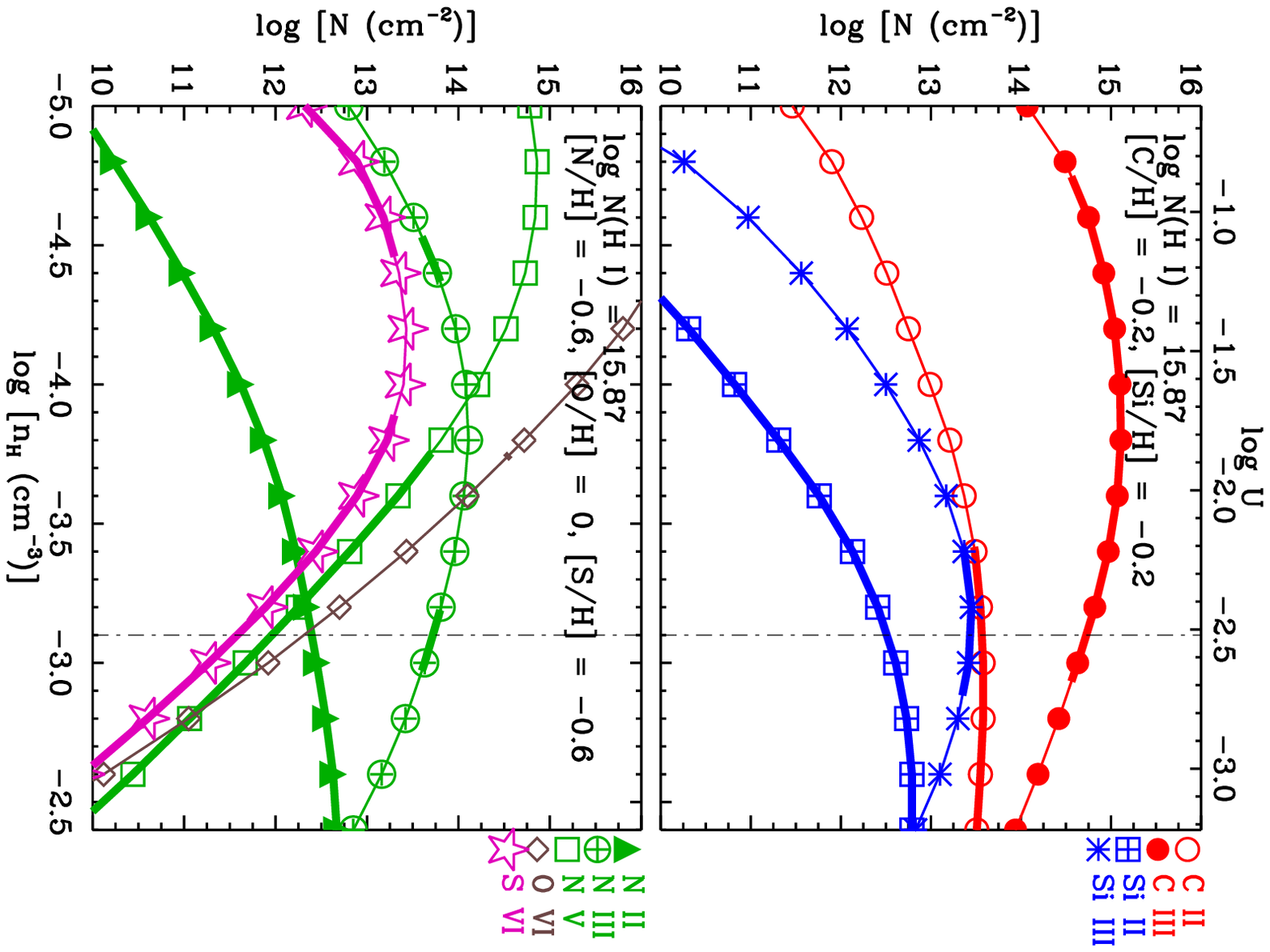}
\includegraphics[totalheight=0.24\textheight, trim=0cm 0cm 0cm 0cm, clip=true, angle=90]{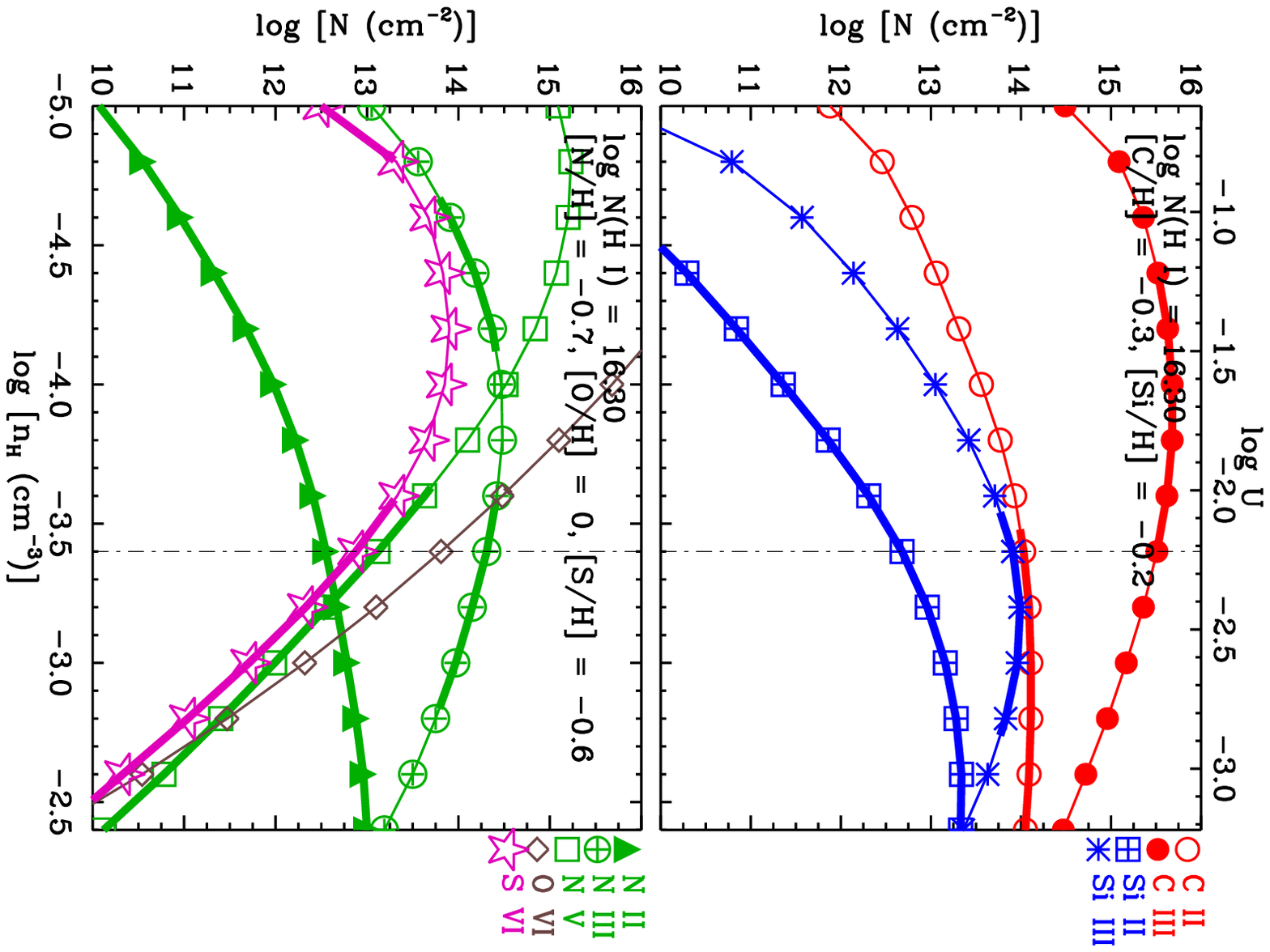}
\caption{Ionic column density predictions from photoionization equilibrium models for the $v = -122$~{\kms} (left top and bottom panels), $-61$~{\kms} (middle panels), $-10$~{\kms} (right panels) components of the $z = 0.41614$ absorber. For clarity, ionic column density predictions are split into top and bottom panels for each of the three components. The Cloudy models for each component were generated for the respective {\HI} column densities obtained from profile fitting the Lyman series lines. The bottom axis is hydrogen number density, the top axis is the ionization parameter defined as $\log U = \log n_{\gamma} - \log n_{\H}$, where $n_{\gamma}$ is the photon number density at energies greater than 1 Ryd. The ionization is from the extragalactic UV background modeled by \citet{KS15b}. The thick portion of the curves, for each ion, mark the $1\sigma$ boundary for the observed column density. For {\CIII} and {\SiIII}, the uncertainties in column density are comparitively larger due to significant line saturation. The {\OVI} is plotted here for reference. The difference in the velocity sub-structure of {\OVI} from the rest of the ions suggest a different origin for {\OVI}, which is discussed in Sec~\ref{sec4.2}}
\label{fig6}
\end{figure*}

It has been pointed out that \citet[hereafter HM12]{HM12} ultraviolet background's estimate of the hydrogen ionizing rate at low-$z$ is lower by a factor of $\sim 2 - 5$ \citep{kollmeier14,shull15,wakker15}. \citet[hereafter KS15]{KS15b} show that this discrepancy is resolved by incorporating in the synthesis of the background spectra, the recent measurements of quasar luminosity function \citep{croom09,palanque13} and star formation rate densities \citep{KS15a}. Figure~\ref{fig5} compares the specific intensity of the HM12 background at $z = 0.4$ with the KS15 model for different escape fractions of Lyman continuum photons. The equilibrium fractions of high ionization species like {\OVI} will be primarily affected by the factor of two increase in the QSO emissivity, whereas the low and intermediate ions will be affected by the enhacement in both galaxy and QSO emissivities. In our photoionization calculations, we use the KS15 ultraviolet background with $4$\% escape fraction of hydrogen ionizing photons to be consistent with the {\HI} photoionization rate estimates of \citet{kollmeier14}. 

The photoionization model predictions for the $v \sim -122$~{\kms} component are shown in Figure~\ref{fig6} (left panel). This feature appears adequately resolved and has, among the three components, well determined {\HI}, {\CIII} and {\SiIII} column densities that are not much affected by saturation. For the observed column density of $\log~N(\HI) = 15.82$, the photoionization curves of Figure~\ref{fig6} show the column density predictions for the different ions at various gas densities. The Si abundance in this component is constrained to [Si/H] $\gtrsim -0.2$, below which {\SiIII} will be underproduced for all densities. For this lower limit on abundance, the observed {\SiIII} and the non-detection of {\SiII} are explained by the models for $n_{\H} = (0.4 - 1.3)~\times~10^{-3}$~{\cc}. A single phase model that is consistent with the observed {\CIII}, {\NIII} and the upper limits on {\CII}, {\NII} is possible at $n_{\H} \sim 0.8 \times 10^{-3}$~{\cc} for [C/H] $= -0.3$ and [N/H] $= -0.6$. The statistical errors in the measured column densities of the metal lines and {\HI} result in a ${\pm}~0.2$~dex error to the estimated abundances. For this moderately ionized gas phase, the model predicts a total hydrogen column density of $\log N(\H) = 18.5$, a line of sight thickness of $L = 1.3$~kpc, and $T = 1.1 \times 10^4$~K. 

The photoionization curves for the $v \sim -67$~{\kms} component are shown in Figure~\ref{fig6} (middle panel). The observed {\SiIII} for this component is valid for [Si/H] $\gtrsim -0.2$, and $n_{\H} = (0.4 - 1.6) \times 10^{-3}$~{\cc}. The unsaturated {\CII} provides an estimate for the carbon abundance in this component. For [C/H] $\lesssim -0.2$, {\CII} becomes a non-detection. Thus [C/H] $\gtrsim -0.2$ for $n_{\H} \geq 0.4 \times 10^{-3}$~{\cc}, densities at which the {\CII} ionization fraction has a maximum. The predicted column densities for the various low and intermediate ions are within the permissible range for $n_{\H} \sim 0.8 \times 10^{-3}$~{\cc} as shown in Figure~\ref{fig6} (middle panel). This model also predicts a $\log N(\H) = 18.6$, $L = 1.6$~kpc, and $T = 1.1 \times 10^4$~K, similar to what we obtain for the $v \sim -122$~{\kms} component. The uncertainties in the measured column densities contribute a $0.6$~dex, $0.4$~dex and $0.3$~dex uncertainty to the derived abundances of C, N and Si respectively.  

For the component at $v \sim -10$~{\kms}, the abundances for C and Si can be derived using {\CII} and {\SiIII}, but with significant uncertainities of $\sim 0.5$~dex coming from the statistical uncertainty in the corresponding {\HI} column density. At [Si/H] $= -0.2$, {\SiIII} is explained for $n_{\H} = (0.3 - 2) \times 10^{-3}$~{\cc} and {\NIII} for [N/H] $= -0.7$ with nearly the same density range. The model predicted {\CII} is consistent with the observed value for [C/H] = $-0.3$ and $n_{\H} \geq 0.3 \times 10^{-3}$~{\cc}. Based on these abundance constrains, a single phase solution for the low and intermediate ions is one where $n_{\H} = 0.4 \times 10^{-3}$~{\cc}, corresponding to $\log N(\H) = 19.4$, $L = 18.9$~kpc, and $T = 1.2 \times 10^4$~K (see Figure~\ref{fig6}, right top and bottom panels). 

The difference in velocity sub-structure of {\OVI} with what is seen for {\CIII}, {\SiIII} and {\HI} clearly precludes the possibility that the {\OVI} ion is tracing this photoionized gas. The {\OVI} prediction for $n_{\H} \sim 0.4 \times 10^{-3}$~{\cc} of the photoionized gas phase is $\sim 0.8$~dex lower than the observed value even for solar [O/H]. Furthermore, the {\OVI} from photoionization has a steeply declining dependency with $\log~n_{\H}$ (e.g., see Figure~\ref{fig6}). This means that for sub-solar oxygen relative abundances, the {\OVI} contribution from this photoionized phase will also come down significantly. The {\OVI} clearly favors an origin in a separate gas phase, which we discuss next.

\subsection{The Origin of {\OVI} Absorption}\label{sec4.2}

The differences between the component structure of {\OVI} when compared to the low, intermediate metal species and the core {\HI} absorption provide the strongest indication that the {\OVI} could be from a phase other than the $T \sim 10^4$~K photoionized gas. The {\OVI} is consistent with having the same origin as the BLA discussed in Sec~\ref{sec3.1.1}. The temperature of this gas phase follows from the large line width of {\HI} compared to {\OVI}. The separate $b$-values of $b(\HI) = 115^{+17}_{-19}$~{\kms} and $b(\OVI) = 48~{\pm}~2$~{\kms} imply a temperature of $T = 7.1_{-2.6}^{+2.7} \times 10^5$~K, with thermal broadening of $b_t(\HI) \sim 108$~{\kms} and $b_t(\OVI) \sim 40$~{\kms}. Whereas for the {\OVI}, the thermal and non-thermal contributions to line broadening are almost equal, for the BLA, $94\%$ of the line broadening is due to the high temperature of the gas. At these high temperatures, the ionization will be dominated by collisions. 

\begin{figure}
\centering
\includegraphics[totalheight=0.35\textheight, trim=0cm 0cm 0cm 0cm, clip=true, angle=90]{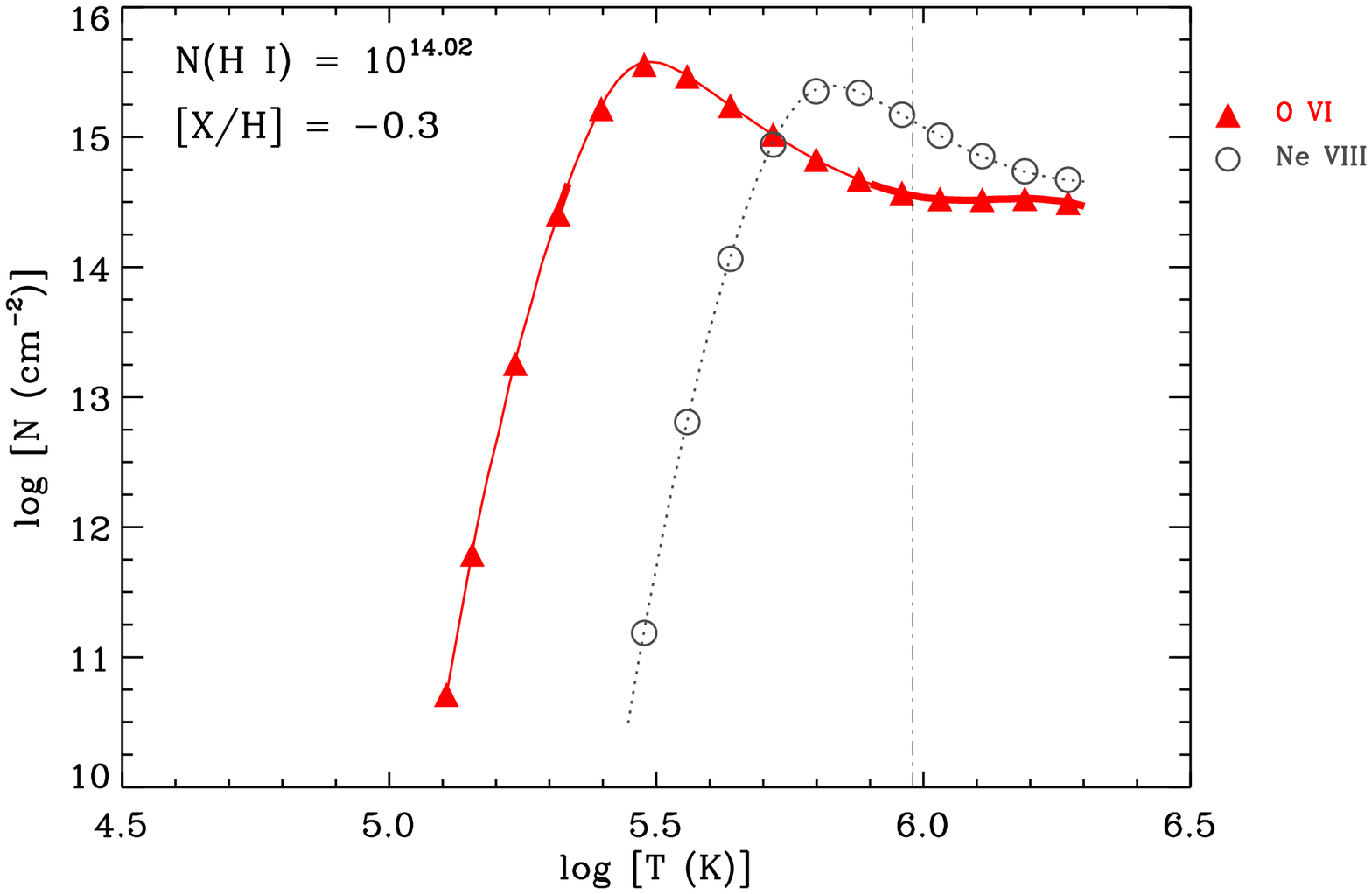}
\caption{\citet{gnat07} CIE model for the {\HI} column density measured for the BLA in the $z = 0.41614$ absorber. The vertical dash-dotted line is at $T = 7.1 \times 10^5$~K, obtained from solving for the temperature using the $b(\HI)$ from the BLA and $b(\OVI)$. The [O/H] $ = -0.3$~dex is set by the need to recover the observed $N(\OVI)$ at the given temperature of the gas. The {\NeVIII} is shown just for comparision, as it is a frequently sought after tracer of warm-hot gas. Both oxygen and neon are modeled for half solar abundance.}
\label{fig7}
\end{figure}

Figure~\ref{fig7} shows \citet{gnat07} collisional ionization equilibrium (CIE) model predictions for {\OVI} at various kinetic temperatures. The column density of {\HI} in these models is the BLA column density of $\log N(\HI) = 14.02$. For the CIE models to match the observed {\OVI} at $T = 7.1 \times 10^5$~K, the oxygen abundance has to be [O/H] $= -0.3~{\pm}~0.2$~dex. For higher metallicity, {\OVI} is overproduced for this equilibrium temperature, and for lower metallicity it is underproduced. However, as we describe in the next paragraph, models that simultaneously consider photoionization and collisional ionization scenarios favor lower oxygen abundance considering realistic values for the size of the warm gas. 

At $T = 7.1 \times 10^5$~K, the CIE conditions predict a very low neutral hydrogen ionization fraction of $f(\HI) = N(\HI)/N(\H) = 3.7 \times 10^{-7}$. This suggests a total hydrogen column density of $N(\H) = 2.8 \times 10^{20}$~{\cmsq}, which is a factor of $\sim 10$ more than the total hydrogen column density in the cooler photoionized gas. At $T > 3 \times 10^{5}$~K, the recombination and cooling rates are comparable, with no differences between the equilibrium and non-equilibrium collisional ionization fraction predictions for metals \citep{sutherland93}. The large baryon content in the {\OVI} gas phase, compared to the strongly absorbing neutral gas, supports the view held by \citet{fox13} and \citet{lehner13} that a proper accounting of the warm {\OVI} phase can double the contribution of such partial/Lyman limit systems ($\log~N(\HI) \sim 16.1 - 16.7$~dex) towards the cosmic baryon budget. 

Even when the ionization in the gas is dominated by collisions, the presence of extragalactic background photons cannot be overlooked. By including the UV ionizing background while keeping the temperature of the gas fixed to $T = 7.1 \times 10^5$~K in Cloudy, we generated hybrid models that simultaneously allow for collisional and photoionization reactions. The hybrid models show that for $n_{\H} \gtrsim 10^{-5}$~{\cc}, the ionization fractions of the various elements are predominantly controlled by collisions. Above that limit, the {\OVI} ionization fraction shows only a weak dependency on density. For densities below this limit, photoionization begins to alter the ion fractions from the pure collisional predictions. At $n_{\H} \sim 10^{-5}$~{\cc}, the hybrid models require a thickness of $\sim 4$~Mpc for the absorber, an exceedingly large value which is difficult to reconcile with the kinematically simple, single component absorption profile of {\OVI} and the BLA. Moreover, absorption over a path length of $4$~Mpc would result in a line broadening of $\sim 280$~{\kms} if the absorber is not decoupled from the Hubble flow, which is inconsistent with the measured width of the lines. Thus, very low densities of $n_{\H} \lesssim 10^{-5}$~{\cc} for the BLA - {\OVI} gas phase can be ruled out. The premise that the absorbing cloud should have a realistic size serves as a constraint on the [O/H] in the hybrid model. For [O/H] $ \lesssim -0.6$~dex, the observed $N(\OVI)$ is produced for $n_{\H} \geq 3 \times 10^{-4}$~{\cc} corresponding to $N(\H) \leq 2.5 \times 10^{20}$~{\cmsq} and $L \leq 260$~kpc. The $1\sigma$ range for the temperature of this warm gas and the associated BLA {\HI} column density, result in an uncertainty in the oxygen abundance of $\sim 0.2$~dex. But the conclusion from CIE, that the BLA - {\OVI} phase possesses a factor of $\sim 10$ more baryons than the moderately ionized gas, is valid for the hybrid ionization scenario as well. 

\begin{figure}
\centering
\includegraphics[totalheight=0.35\textheight, trim=0cm 0cm -1cm 0cm, clip=true, angle=90]{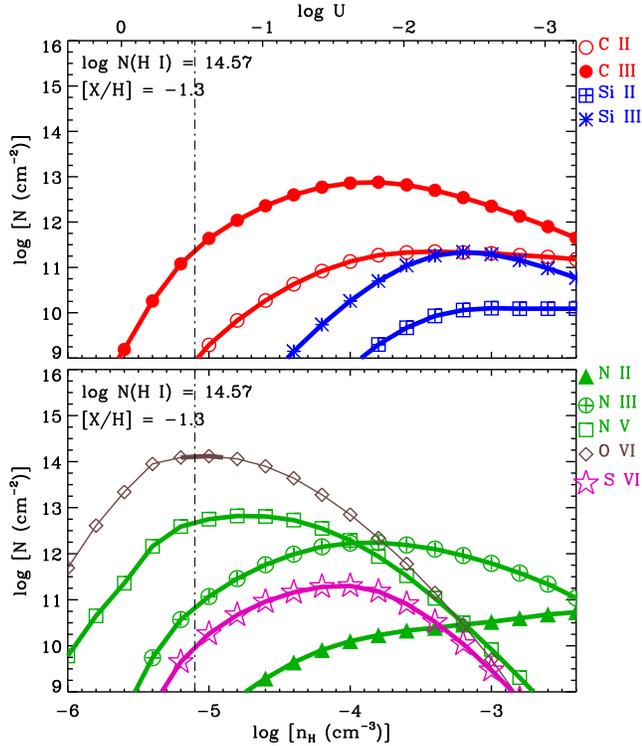}
\caption{Column density predictions from photoionization equilibrium models for the $v = -3$~{\kms} {\HI} - {\OVI} absorber at $z = 0.41950$. For clarity, the model predictions for the different ions are split into top and bottom panels. The thick portion of the curves mark the $1\sigma$ boundary for the observed column density. Except {\OVI}, all other metal lines are non-detections. The dash-dot line indicates the ionization parameter at which {\OVI} is reproduced by the models simultaneously being consistent with the upper limits for the rest of metal ions.}
\label{fig8}
\end{figure}

We note that the {\NeVIII} from the collisionally ionized warm gas will be stronger than {\OVI} at the predicted temperature of $T = 7.1 \times 10^5$~K, for [Ne/H] $\sim$ [O/H]. The absorber could have been a candidate for {\NeVIII} detection, but the incidence of a Lyman limit absorber at $z = 0.31$ leaves little flux at $\lambda < 1200$~{\AA}. The properties of the warm gas, nevertheless, are consistent with the general physical state of {\NeVIII} absorbers given by both observations and simulations \citep{tepper13}.

\section{Ionization \& Abundances in the $z = 0.41950$ Absorber}\label{sec5}

The kinetic temperature $T = 4.6^{+0.3}_{-0.2} \times 10^4$~K, measured by the narrow {\HI} and {\OVI} line widths is consistent with temperature of the gas being driven by photoionization. At these temperatures, electron impacts will not contribute significantly to the ionization of {\HI} or metals and hence collisional ionization can be neglected. Figure~\ref{fig8} shows the photoionization models for the narrow {\HI} - {\OVI} cloud when irradiated by the KS15 ionizing spectrum. The observed $N(\OVI)$ allows us to impose a lower limit on the oxygen abundance. For [O/H] $\leq -1.3$~dex, {\OVI} is underproduced for all densities. The photoionization models shown in Figure~\ref{fig8} are for a metallicity equal to this lower limit for oxygen. At that limiting abundance, the required {\OVI} is produced for $n_{\H} = 7.9 \times 10^{-6}$~{\cc}. Such low densities would result in large path lengths of $L = 2.5$~Mpc. Over a megaparsec scale it is common to expect discrete velocity and ionization substructures in the absorbing cloud. Given the simple Gaussian like optical depth profile of {\OVI} and the narrow {\HI}, it is unlikely that the absorption is happening across such vast length scales. If the absorption is from a structure that has not decoupled from the Hubble flow, then we expect a line broadening due to Hubble expansion of $v = H(z)L = 215$~{\kms}, which is significantly larger than the {\HI} or {\OVI} line widths. The {\OVI} can come from clouds of smaller size, provided the densities are higher. As the photoionization curves show, {\OVI} has a steep dependency on density. Hence higher density photoionization solutions would also require higher oxygen abundances. For example, at [O/H] $\sim -1.0$~dex, the observed $N(\OVI)$ is recovered from $n_{\H} = 2.5 \times 10^{-5}$~{\cc}, tracing a total baryonic column of $N(\H) = 1.3 \times 10^{19}$~{\cmsq} over a significantly smaller cloud thickness of $L = 129$~kpc. The above gas phase is consistent with the upper limits from the non-detection of {\NII}, {\NIII}, {\NV}, {\SiII}, {\SiIII} and {\SVI} for the 1/10th solar metallicity. The photoionization predicted temperature of $T = 4.1 \times 10^4$~K also agrees with the line widths of {\HI} and {\OVI}. Such higher density solutions require departures from solar [C/O] if the models are to be consistent with the absence of {\CIII} from the same gas phase. For example at a metallicity of $-0.5$~dex, the [C/O] $\sim -0.3$~dex for the {\CIII} to be a non-detection. Significant variation in [C/O] abundance would require a different nucleosynthesis enrichment history, as both C and O are primary elements synthesized by massive stars. Such lower abundances of C to O are seen for higher column density {\HI} systems like DLAs and LLSs \citep{lehner13,dutta14}. Assuming [C/O] of solar, the constraint on oxygen abundance will be $-0.7 \lesssim$~[O/H]~$\lesssim -1.3$~dex.  

The ionization and physical conditions in the $v \sim -72$~{\kms} cloud is less precisely determined given the absence of any metal lines to go with the broad {\HI}. Photoionization models for the estimated column density of $\log[N(\HI)] = 13.48$ show that abundances [X/H] $\leq 0$ and densities of $n_{\H} < 5 \times 10^{-5}$~{\cc} satisfy the upper limits on the metal ion column densities. The model predicted gas temperature of $T = 2 \times 10^4$~K would imply that $\sim 70$\% of the {\HI} line width is due to non-thermal broadening. 

Alternatively, if $b(\HI) = 58~{\pm}~6$~{\kms} is purely due to thermal broadening, the corresponding temperature will be in the range $T = (1.7 - 2.5)~\times~10^5$~K. Contributions from turbulence in the gas, along with Hubble broadening ($b_{\mathrm{Hubble}} < 30$~{\kms} typically for $\sim$~kpc structures, Valageas, Schaeffer, \& Silk 2002) would result in lower temperatures. Given the lack of information to resolve the mechanisms contributing towards line broadening, the temperature estimate can be considered as an upper limit. At $T = 2 \times 10^5$~K, if collisional processes are dominating the ionization in the gas, then the total baryon column density in the broad component will be $N(\H) = 3 \times 10^{19}$~{\cmsq}, assuming CIE fractions \citep{gnat07}. The total gas content in this shallow and broad {\HI} component thus comes out as an order of magnitude more than the amount of baryons present in the kinematically adjacent photoionized gas phase of the absorber. At $T = 2 \times 10^5$~K, the column density limit from non-detection of $\log~N(\OVI) < 13.3$~dex, places an upper limit of [O/H] $< -0.8$~dex in the collisionally ionized gas.

\begin{table*} 
\caption{\textsc{Galaxies in the vicinity of the absorbers}}
\begin{center}
\small
\begin{tabular}{lcccccccc}
\hline
R.A.  &  Dec.   &   $z_{gal}$   &  $\Delta v$ (\kms)  &  $\eta$ (arcmin)   &   $\rho$  (Mpc)  &  $g$ (mag)  &   M$_g$ & ($L/L^*$)$_g$ \\
\hline
\hline
\multicolumn{8}{c}{$z_{abs} = 0.41614$ ($v = 0$~{\kms}) Absorber} \\ \hline \hline
$150.14156$ &  $59.72934$ & $0.41437$ & $-376~{\pm}~21$ & $3.6$ & $1.2$ & $20.97~{\pm}~0.09$ & -22.34 & 2.8 \\
$150.42517$ &  $59.73372$ & $0.41303$ & $-659~{\pm}~16$ & $5.0$ & $1.7$ & $20.49~{\pm}~0.06$ & -22.82 & 4.3 \\
$150.43219$ &  $59.81284$ & $0.42062$ & $+947~{\pm}~15$ & $6.9$ & $2.3$ & $21.04~{\pm}~0.07$ & -22.40 & 2.9 \\
$150.15218$ &  $59.61503$ & $0.41422$ & $-406~{\pm}~17$ & $8.0$ & $2.7$ & $20.76~{\pm}~0.10$ & -22.06 & 2.1 \\
$150.20506$ &  $59.43657$ & $0.41189$ & $-901~{\pm}~18$ & $18.1$ & $6.0$ & $20.85~{\pm}~0.09$ & -22.59 & 3.5 \\
$150.88540$ &  $60.04693$ & $0.41456$ & $-334~{\pm}~21$ & $26.4$ & $8.8$ & $20.77~{\pm}~0.06$ & -22.56 & 3.4 \\
$149.66518$ &  $60.11736$ & $0.41496$ & $-249~{\pm}~17$ & $28.9$ & $9.6$ & $20.97~{\pm}~0.05$ & -22.32 & 2.7 \\
\hline
\multicolumn{8}{c}{$z_{abs} = 0.41950$ ($v = +710$~{\kms}) Absorber} \\ \hline \hline
$150.43219$ &  $59.81285$ & $0.42062$ & $+236~{\pm}~14$ & $6.9$ & $2.3$ & $21.04~{\pm}~0.07$ & -22.40 & 2.9 \\
$149.66518$ &  $60.11736$ & $0.41496$ & $-959~{\pm}~16$ & $29.0$ & $9.7$ & $20.97~{\pm}~0.06$ & -22.32 & 2.7 \\
\hline
\end{tabular}
\label{tab4}
\end{center}
\scriptsize{Comments. - Galaxies within 30 arcminutes of projected separation and within $|\Delta v| = 1000$~{\kms} of the absorbers. The $z$ values are SDSS spectroscopic redshifts. $\Delta v$ correspond to the systemic velocities of the galaxies with respect to the absorber. The error in velocity separation comes from the uncertainty in the spectroscopic redshift. The projected separation $\rho$ was calculated from the angular separation assuming a $\Lambda$CDM universe with parameters of $H_0 = 69.6$~{\kms}~Mpc$^{-1}$, $\Omega_m = 0.286$, $\Omega_{\Lambda} = 0.714$ \citep{bennett14}. The galaxy absolute magnitudes were calculated using the distance modulus expression, with the luminosity distances estimated for each $z_{gal}$ for a $\Lambda CDM$ universe \citep{wright06}. Appropriate $K$ corrections were applied using the analytical expression given by \citet{chilingarian10}. The Schechter absolute magnitude $M^*_g = -21.237$ for $z = 0.4$ was taken from \citet{ilbert05}. The distribution of galaxies near the $z = 0.41614$ absorber is shown in Figure~\ref{fig9}.}
\end{table*}


Hybrid models suggest that for {\OVI} to be a non-detection at $T = 2 \times 10^5$~K, the oxygen abundance in the BLA gas phase has to be [O/H] $< -0.7$, which is consistent with the corresponding upper limit in the kinematically adjacent {\HI} - {\OVI} cloud. For [O/H] $< -0.7$, density is $n_{\H} \lesssim 3 \times 10^{-3}$~{\cc}, $N(\H) > 10^{19}$~{\cmsq} and $L > 1$~kpc. 

\section{Galaxies Near the Absorber}\label{sec6}

\begin{figure}
\centering
\includegraphics[totalheight=0.3\textheight, trim=0cm 0cm 0cm 0cm, clip=true]{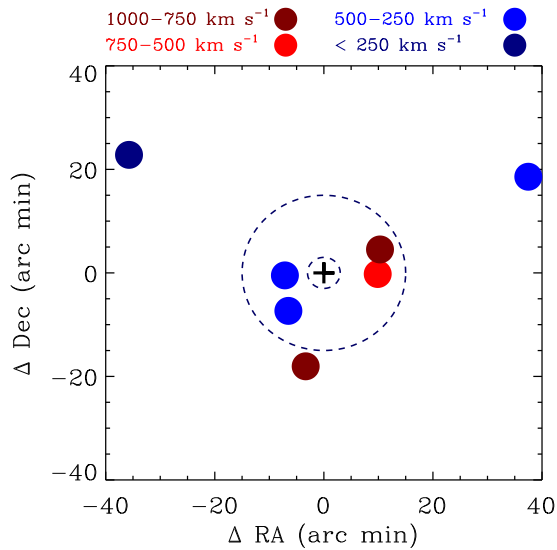}
\caption{The distribution of galaxies with SDSS spectroscopic redshifts that place them within $|\Delta v| = 1000$~{\kms} and 30 arcminute from the $z = 0.41614$ absorber. The "+" sign corresponds to the line of sight towards the background quasar. The two concentric dashed circles indicate uniform projected separations of $~1$~Mpc and $5$~Mpc respectively from the line of sight. Further information on these galaxies is given in Table~\ref{tab4}. The multiphase absorber seems to be residing in a galaxy group environment.}
\label{fig9}
\end{figure}

The SBS0957+599 sightline is in the SDSS footprint. We searched the DR12 SDSS database for galaxies close to the absorbers. Within $|\Delta v| =1000$~{\kms} velocity separation and $30^{\prime}~\times~30^{\prime}$ ($\sim 10$~Mpc) projected separation of the $z = 0.41614$ absorber are seven galaxies with SDSS spectroscopic redshifts. The galaxy distribution at the location of the absorber is shown in Figure~\ref{fig9} and their information summarized in Table~\ref{tab4}. 

SDSS is sampling only the brightest galaxies at $z \sim 0.4$, as evident from the $> 2~L*$ luminosities we estimate for all the galaxies in Table~\ref{tab4}. The survey's $90$\% spectroscopic completeness limit of $r < 17.8$ \citep{strauss02} corresponds to $\gtrsim 3~L*$ at $z \sim 0.4$ \citep{ilbert05}.  The nearest galaxy seen by SDSS is at a projected separation of $1.2$~Mpc from the absorber. This $2.8~L*$ galaxy has an extended morphology and an emission line spectrum. The flux at H$\alpha$ is $f_{\H\alpha} = 14.3 \times 10^{-17}$~erg~cm$^{-2}$~s$^{-1}$. Using the conversion factor given by \citet{kennicutt94}, we estimate a star formation rate of SFR$(\H\alpha) = 0.7$~M$_\odot$~yr$^{-1}$, typical of normal galaxies. The scaling relation $R_{\mathrm{vir}} = 250~(L/L*)^{0.2}$~kpc given by \citet{prochaska11b} yields a virial radius of $R_{\mathrm{vir}} \sim 310$~kpc for this galaxy. The separation of the galaxy from the line of sight is a factor of $\sim 4$ larger than this. Given this wide separation and the low SFR, it is unlikely that the absorber has an origin in gas bound to this galaxy. 

The impact parameters of the other galaxies range from $\sim 3 - 10$~Mpc. Because of the incompleteness of the galaxy sample for sub-$L*$ luminosities, and the small number of galaxies identified with redshifts similar to that of the absorber, we do not adopt any standard algorithm to formally define a galaxy group in this region. However, we note that the four galaxies within $|\Delta v| < 500$~{\kms} of the absorber have a narrow velocity dispersion of $\sim 59$~{\kms} and an average velocity of $\sim - 342$~ {\kms} with reference to the absorber. Such offsets in velocity, also seen between several warm absorbers and galaxy groups studied by \citet{stocke14}, could be related to the general kinematic nature of warm gas near galaxy groups. 

Including the photometric redshifts from the SDSS reveals four additional galaxies within $\rho = 5$~Mpc projected distance of the absorber, with the nearest galaxy at $\sim 1.9$~Mpc. The errors associated with the photometric redshifts are $\delta z/z \sim 0.03 - 0.3$, which allow for uncertainties of $\sim 10^3 - 10^4$~{\kms} in the systemic velocities of these galaxies. Spectroscopic observations of these galaxies can yield more detail on this absorber environment. In any case, the preponderance of galaxies suggests that the environment could be one of galaxy groups, with the absorption tracing gas near to $1$~Mpc of the galaxies revealed by SDSS. 

Two of the galaxies from the above sample are within $\Delta v = 1000$~{\kms} of the $z = 0.41950$ absorber as well, but again at significantly large impact parameters of $> 2$~Mpc (see Table~\ref{tab4}). The galaxy nearest in impact parameter shows an emission line dominated spectrum with $f_{\H\alpha} = 131.1 \times 10^{-17}$~erg~cm$^{-2}$~s$^{-1}$. The corresponding SFR$(\H\alpha) = 6.7$~M$_\odot$~yr$^{-1}$ (Kennicutt {\etal} 1994) indicate that the galaxy has a moderate star-formation rate. Given the velocity offsets between the absorber and the galaxies and the large impact parameters, the likelihood of the line of sight tracing the gaseous envelope of either galaxy is negligible. We note that there are three more galaxies with photometric redshifts within $500$~{\kms} and $\rho < 5$~Mpc of the absorber. The nearest of these galaxies is at $2.2$~Mpc impact parameter. However, the photometric redshift errors signficiantly extend the uncertainty in the velocity offset of these galaxies with the absorber.

\section{On The Origin of the Absorbers}\label{sec7}

Having information on the size of {\OVI} absorbing regions can help in the physical interpretation of individual absorption systems, whether the {\OVI} is tracing virialized galactic scale structures or intergalactic gas. Surveys that explore the association between galaxies and absorbers find a wide spatial distribution for {\OVI} around galaxies compared to low ionization metal lines \citep{prochaska11b,tumlinson11a,stocke14}. An ionization model independent estimate on the size of the warm {\OVI} halo around a galaxy was obtained by \citet{muzahid14}. Using two closely separated lines-of-sight, \citet{muzahid14} estimated a coherence length of $\sim 280$~kpc for the {\OVI} absorption from the circumgalactic medium of a $\sim 1.2~L^*$ galaxy, consistent with the emerging view that sub-$L^*$ and brighter galaxies can have covering fractions of nearly unity for {\OVI} out to their halo virial radii \citep[e.g.,][]{prochaska11b}. 

In regions where multiple galaxies are present, there is an enhancement in the covering fraction of {\OVI}, with detections extending out to impact parameters which are as much as $\sim 3$ times the halo virial radii of the nearest galaxies \citep{johnson15,mathes14}. Interestingly, such an enhancement is not found for the cooler ($T \sim 10^4$~K), dense ($n_{\H} \sim 10^{-3}$~{\cc}) gas probed by {\CII}, {\SiII}, {\MgII} and similar low ionization potential lines, suggesting that these could be parsec-scale clouds embedded within a much extended warm halo \citep{ford14,muzahid14}. This environmental dependence is reflected in the absorbers discussed here. The $z = 0.41614$ absorber, with its mix of low and high ionization gas, is residing within a few Mpc of an overdense region of space where the spread of warm {\OVI} is likely to be wider than halo sizes. Moreover, galaxy interactions, ram-pressure stripping and similar environmental effects are pathways by which low ionization ISM can also be dragged to large separations from galaxies. In the $z = 0.41950$ absorber, we find evidence only for high ionization gas, which is consistent with the relatively low density of luminous galaxies close to the line of sight. 

The galaxies identified by SDSS are more than $1$~Mpc distant from either absorber. As the $z = 0.41614$ absorber is a partial Lyman limit system with $\log~N(\HI) \sim 16.6$~dex, one may wonder whether the magnitude limited observations of SDSS have missed sub-$L^*$ sources closer to this absorber. This remains a distinct possibility in the light of several absorber/galaxy studies which find that {\OVI} bearing gas can be seen near ($\rho \lesssim 800$~kpc) many $L \gtrsim 0.1~L^*$ galaxies \citep{stocke06, wakker09, prochaska11b}. \citet{prochaska11b} conclude that the circumgalactic environments of sub-$L^*$ galaxies ($0.1L^* < L < L^*$) could be predominantly responsible for relatively strong {\OVI} systems with $W_r (1032) > 30$~m{\AA}. At the same time, evidence seems to also suggest that the virialized halos of very faint dwarf galaxies ($L \gtrsim 0.01~L*$) may have a lesser role to play in the incidence of {\OVI} absorbers \citep{tumlinson05, prochaska11b,mathes14}. For example, from studying a sample of 14 galaxies that lie within $\rho \sim 3~R_{vir}$ of background quasars, \citet{mathes14} find a significantly lower frequency of {\OVI} detections around low mass halos ($\sim 10^{11.5}$~M$_{\odot}$), suggesting a fairly large escape fraction of $\sim 90$\% for halo gas, compared to a modest $\sim 35$\% from high mass halos \citep{mathes14}. Thus, low mass galaxies may not retain much of {\OVI} bearing outflows. 
Since the SDSS has only found the brightest galaxies ($\gtrsim 2~L^*$) near the absorber, there remains the possibility that the $z = 0.41614$ {\OVI} - partial Lyman limit absorber could be dynamically linked to the halo of an undetected sub-$L*$ galaxy closer to the absorber. 

Galaxy feedback mechanisms such as correlated supernova driven winds and AGN outflows can influence the temperature and the chemical composition of halos \citep{veilleux05}. At low-$z$, starforming galaxies are known to have more warm {\OVI} in their halos than passive galaxies, implying that the {\OVI} can be a direct result of gas shocked by supernova ejecta \citep{tumlinson11a}. At higher redshifts ($z \sim 3$) also, many strong {\OVI} absorbers ($\log N(\OVI) \gtrsim 14.5$) are found to have velocities, densities and temperatures consistent with outflows \citep{lehner14}. For lines of sight probing active outflows or winds close to the galaxy, the velocity extents of absorption features are generally broad ($\Delta v \gtrsim 400$~{\kms}) and kinematically complex \citep{heckman01,grimes09,tripp11,tumlinson11a,muzahid14}. This is not the case for either of our {\OVI} absorbers. The sub-solar metallicity which we obtain for the low and high ionization gas phases is also not usually found in metal rich outflows \citep{muzahid14}. Thus, there is no definitive evidence for the absorbers discussed in this paper being directly associated with a galactic-scale wind. These warm absorbers can be tracers of ancient outflows (wind age $> 1$~Gyr) which, as the simulations of \citet{ford14} suggest, can end up diffusely distributed at galactocentric distances of $d \gtrsim 500$~kpc where absorption from the ambient IGM also dominates. With the limited SDSS sampling of galaxies discussed in this paper, it is difficult to distinguish an outflow generated {\OVI} from the several other possible scenarios that can give rise to warm {\OVI} in galaxy halos \citep[see ][]{heckman02,sembach03,maller04,tripp08}. 

Based on models of structure formation, it has been hypothesized that galaxy groups can have intragroup gas at $T \sim 10^5 - 10^6$~K, temperatures too low to discrimate their emission from the soft X-ray background \citep{mulchaey96}. Presently, such a medium can be probed only through high ionization absorption lines such as {\OVI}, {\NeVIII} and thermally-broadened {\Lya} against the light from background QSOs. It is important to explore this gas reservoir, as it potentially harbours as much as $\sim 4$\% of the total baryons in the low-$z$ universe, on a par with the fraction of baryons trapped inside galaxies \citep{savage14}. A recent example is given by \citet{stocke14} whose sample of BLA - {\OVI} warm absorbers is morely likely to be from intragroup gas than individual galaxy halos. The warm {\OVI} is hypothesized to arise in conductive interface layers between cold $T \lesssim 3 \times 10^4$~K clouds and a hitherto undetected diffuse $T \sim 10^{6.5}$~K intragroup medium. 

Hydrodynamic simulations also predict that a substantial fraction of baryons are to be found along large scale filaments connecting galaxy overdensity regions. In \citet{narayanan11}, a BLA - {\OVI} absorber is identified as tracing $T \sim 1.4 \times 10^5$~K intergalactic gas along a nearby ($v_{\mathrm{HELIO}} = 3081$~{\kms}) galaxy filament, with several loose groups of galaxies beyond $\rho \sim 1.5$~Mpc of projected separation. Another example of baryonic reservoirs along galaxy filaments is given by \citet{wakker15}. About $85$\% of the {\Lya} systems from their sample of 15 absorbers, which includes BLAs, appear to be tracing intergalactic filament gas that lies far outside of the virial radii of the nearest galaxies that trace the filament. Along similar lines, \citet{tejos16} find a $\sim 4 - 7$~times higher covering fraction for BLAs near galaxy overdensities, compared to random fields. In that study also the warm gas was identified as being away from individual galaxy halos, but possibly along cluster filaments. 

The absorbers studied here are also consistent with an intragroup origin or gas associated with a much larger galaxy filament. The {\OVI} and the associated BLA in the $z = 0.41614$ system can be produced by collisional process at the interface layers between the photoionized gas phase and a hot exterior medium. In the case of the $z = 0.41614$ system, the photoionized gas could be recycled material from earlier epochs of galaxy outflows or interactions within the group environment, as proposed for metal rich ([X/H] $\gtrsim -0.3$) Lyman limit systems \citep{lehner13}.  The {\OVI} in the $z = 0.41950$ system is likely produced in a low density medium that is predominantly photoionized. However, even in this latter case, the BLA could be transition temperature plasma, with metallicity so low to have little {\OVI}. 

\section{SUMMARY \& CONCLUSIONS}\label{sec8}

In this paper, we have analyzed the physical conditions in two intervening multiphase {\OVI} absorbers at $z = 0.41614$ and $z = 0.41950$, separated from each other by $\Delta v = 710$~{\kms}. Both absorbers are detected in the COS spectrum of the background quasar SBS~$0957+599$ and they show clear evidence for the presence of gas with $T \sim 7 \times 10^5$~K and $T \lesssim 2 \times 10^5$~K respectively. Our main conclusions are :

\begin{enumerate}

\item The {\CII}, {\CIII}, {\SiII}, {\SiIII}, {\NII} and {\NIII} in the $z = 0.41614$ system are consistent with cool photoionized clouds of $T \sim 4 \times 10^4$~K having densities of $n_{\H} \sim 0.4 \times 10^{-3}$~{\cc}, total hydrogen column densities of $N(\H) \sim 3 \times 10^{19}$~{\cmsq} with absorption happening over path lengths of $\lesssim 20$~kpc. This gas phase has elemental abundances of [C/H] $\sim -0.3$~dex, [Si/H] $\sim -0.2$~dex, [N/H]~$\lesssim -0.7$~dex and [S/H] $\lesssim -0.6$~dex, with uncertainties of $~{\pm}~0.4$~dex. 

\item There is a BLA component to the {\HI} absorption at $z = 0.41614$ with $\log~N(\HI) = 14.02~{\pm}~0.23$~dex and $b(\HI) = 115^{+17}_{-19}$~{\kms}, revealing the presence of a warm medium. The BLA and {\OVI} are consistent with an origin in the same gas phase. The widths of the {\HI} and {\OVI} lines solve for $T = 7.1 \times 10^5$~K, a temperature too hot for any of the low or intermediate ionization species to survive. Under pure collisional ionization, this warm phase has [O/H] $= -0.3~{\pm}~0.2$~dex and traces a substantial baryonic column of $N(\H) = 2.8 \times 10^{20}$~{\cmsq}. The amount of baryons present in this warm medium is a factor of 10 higher than the cooler photoionized gas where the {\HI} column density is $\sim 2.5$~dex higher. Hybrid models that simultaneously allow for photoionization and collisional ionization favor a lower abundance of [O/H] $\lesssim -0.6$~dex set by constraints on the size of the absorber. 

\item The $z = 0.41950$ absorber is detected only in {\HI} and {\OVI}. The {\Lya} profile shows the presence of a shallow, broad component and a stronger, narrow component offset from each other by $\sim 70$~{\kms}. The {\OVI} does not show any evidence for sub-structure and is kinematically centered on the narrow {\HI} component. The $b(\HI)$ and $b(\OVI)$ indicates $T = 4.6^{+0.3}_{-0.2} \times 10^4$~K where collisional ionization is not important. The {\OVI} permits photoionization at low-densities of $n_{\H} \sim 10^{-5}$~{\cc}, $-0.7 \lesssim $~[O/H]~$\lesssim -1.3$~dex, baryonic column densities of $N(\H) \sim 10^{19}$~{\cmsq} and line-of-sight thickness of $L \sim 130$~kpc. 

\item The BLA in the $z = 0.41950$ absorber has a $b(\HI) = 58~{\pm}~6$~{\kms} suggesting $T = (1.7 - 2.5) \times 10^5$~K for pure thermal broadening. The temperature supports collisonal ionization in gas with baryonic column densities of $N(\H) \lesssim 3 \times 10^{19}$~{\cmsq}. The oxygen abundance in this warm gas phase is constrained (from the non-detection of coincident {\OVI}) to [O/H] $< -0.8$~dex, if collisions are dominating the ionization in this medium. 

\item The SDSS database shows four spectroscopically confirmed galaxies with $L \gtrsim 2L^*$ within $30$~arcminute and $|\Delta v| = 500$~{\kms} velocity of the $z = 0.41614$ absorber. At the redshift of the absorber, the SDSS is 90\% complete for $L \gtrsim 3L^*$. The nearest galaxy is at a projected separation of $1.2$~Mpc. The narrow velocity dispersion of $\sim 59$~{\kms} between the four galaxies indicate that the absorber is tracing gas associated with a galaxy overdensity environment. There are three additional spectroscopically confirmed galaxies within $1000$~{\kms} of this absorber. The nearest luminous source to the $z = 0.41950$ absorber is a moderately star-forming disk galaxy whose systemic redshift places it at $235$~{\kms} and $2.3$~Mpc impact parameter. Both {\OVI} absorbers are possibly tracing out multi-phase intragroup gas, or the gaseous envelope of a closer-by sub-$L^*$ galaxy undetected by the SDSS. 

\item Analysis of the two absorbers highlights the diverse ionization conditions as well as the physical environment in which {\OVI} absorption arises. The metal species cannot be used as a blind tracer of $T \sim 10^5 - 10^6$~K gas. In multi-phase absorbers such as those presented here, the {\OVI} data has to be interpreted in the light of additional information from thermally broad {\Lya} or more highly ionized species like {\NeVIII} or {\MgX} to discriminate between photoionization and collisional ionization processes.  

\end{enumerate}

\section{Acknowledgments}\label{sec9}

We thank the referee, John Stocke, for a thorough and rapid review of the manuscript, and for providing several valuable suggestions for improving the impact and presentation of this work. SP \& AN thank IIST and the Department of Space, Government of India for the financial support. 

\def\aj{AJ}%
\def\actaa{Acta Astron.}%
\def\araa{ARA\&A}%
\def\apj{ApJ}%
\def\apjl{ApJ}%
\def\apjs{ApJS}%
\def\ao{Appl.~Opt.}%
\def\apss{Ap\&SS}%
\def\aap{A\&A}%
\def\aapr{A\&A~Rev.}%
\def\aaps{A\&AS}%
\def\azh{AZh}%
\def\baas{BAAS}%
\def\bac{Bull. astr. Inst. Czechosl.}%
\def\caa{Chinese Astron. Astrophys.}%
\def\cjaa{Chinese J. Astron. Astrophys.}%
\def\icarus{Icarus}%
\def\jcap{J. Cosmology Astropart. Phys.}%
\def\jrasc{JRASC}%
\def\mnras{MNRAS}%
\def\memras{MmRAS}%
\def\na{New A}%
\def\nar{New A Rev.}%
\def\pasa{PASA}%
\def\pra{Phys.~Rev.~A}%
\def\prb{Phys.~Rev.~B}%
\def\prc{Phys.~Rev.~C}%
\def\prd{Phys.~Rev.~D}%
\def\pre{Phys.~Rev.~E}%
\def\prl{Phys.~Rev.~Lett.}%
\def\pasp{PASP}%
\def\pasj{PASJ}%
\def\qjras{QJRAS}%
\def\rmxaa{Rev. Mexicana Astron. Astrofis.}%
\def\skytel{S\&T}%
\def\solphys{Sol.~Phys.}%
\def\sovast{Soviet~Ast.}%
\def\ssr{Space~Sci.~Rev.}%
\def\zap{ZAp}%
\def\nat{Nature}%
\def\iaucirc{IAU~Circ.}%
\def\aplett{Astrophys.~Lett.}%
\def\apspr{Astrophys.~Space~Phys.~Res.}%
\def\bain{Bull.~Astron.~Inst.~Netherlands}%
\def\fcp{Fund.~Cosmic~Phys.}%
\def\gca{Geochim.~Cosmochim.~Acta}%
\def\grl{Geophys.~Res.~Lett.}%
\def\jcp{J.~Chem.~Phys.}%
\def\jgr{J.~Geophys.~Res.}%
\def\jqsrt{J.~Quant.~Spec.~Radiat.~Transf.}%
\def\memsai{Mem.~Soc.~Astron.~Italiana}%
\def\nphysa{Nucl.~Phys.~A}%
\def\physrep{Phys.~Rep.}%
\def\physscr{Phys.~Scr}%
\def\planss{Planet.~Space~Sci.}%
\def\procspie{Proc.~SPIE}%
\let\astap=\aap
\let\apjlett=\apjl
\let\apjsupp=\apjs
\let\applopt=\ao
\bibliographystyle{mn}
\bibliography{mybib}

\label{lastpage}

\end{document}